\documentclass[structabstract]{aa}  
  
\usepackage{graphicx}
\usepackage{txfonts}
\begin{document}
\titlerunning{HS\ 2236+1344}
\title{On the properties of the interstellar medium in extremely metal-poor blue compact dwarf galaxies:}
\subtitle{GMOS-IFU spectroscopy and SDSS photometry of the double-knot galaxy HS\ 2236+1344}

   \author{Lagos, P.,
          \inst{1}
          Papaderos, P.,
          \inst{1}
          Gomes, J. M.,
          \inst{1}
          Smith Castelli, A. V.,
          \inst{2,3}
          \and
          Vega, L. R.
          \inst{4}
          }

   \institute{Centro de Astrof\'{\i}sica da Universidade do Porto, Rua das Estrelas, 4150-762 Porto, Portugal\\
   \email{plagos@astro.up.pt}
         \and 
              Instituto de Astrof\'{\i}sica de La Plata (CCT La Plata, CONICET, UNLP), Paseo del Bosque, B1900FWA 
              La Plata, Argentina
         \and     
              Facultad de Ciencias Astron\'omicas y Geof\'{\i}sicas, Universidad Nacional de La Plata, Paseo del Bosque, 
              B1900FWA La Plata, Argentina
         \and
              Observatorio Astron\'omico de C\'ordoba, Laprida 854, C\'ordoba, 5000, Argentina
             }

%   \date{Received September 15, 1996; accepted March 16, 1997}
% \date{\textbf{version 5}}

  \abstract
{}
{
The main goal of this study is to carry out a spatially resolved investigation
of the warm interstellar medium (ISM) in the extremely metal-poor Blue Compact
Dwarf (BCD) galaxy  \object{HS\ 2236+1344}. 
Special emphasis is laid on the analysis of the spatial distribution of chemical abundances, 
emission-line ratios and kinematics of the ISM, and to the recent star-forming (SF) activity in this galaxy. 
}
{
This study is based on optical integral field unit spectroscopy data from Gemini Multi-Object Spectrograph 
(GMOS) at the Gemini North telescope and archival Sloan Digital Sky Survey (SDSS) images. 
The galaxy has been observed at medium spectral resolution ($R\sim 4000$) over the spectral range 
from $\sim$ 4300 $\rm \AA$ to 6800 $\rm \AA$. The data were obtained in two different positions
across the galaxy, obtaining a total 4\arcsec$\times$8\arcsec\ field
which encompasses most of its ISM. 
}
{Emission-line maps and broad-band images obtained in this study indicate that HS\ 2236+1344 hosts
three Giant H\,{\sc ii} regions (GH\,{\sc ii}Rs).
Our data also reveal some faint curved features in the BCD periphery that
might be due to tidal perturbations or expanding ionized-gas shells.
The ISM velocity field shows systematic gradients along the major axis of the
BCD, with its south-eastern and north-western half differing by $\sim$80 km/s in their recessional velocity. 
The H$\alpha$ and H$\beta$ equivalent width distribution in the central part of
HS\ 2236+1344 is consistent with a very young ($\sim$3 Myr)
burst. Our surface photometry analysis indicates that the ongoing starburst 
provides $\sim$50\% of the total optical emission, similar to other BCDs.
It also reveals an underlying lower-surface brightness component
with moderately red colors, which suggest that the galaxy has undergone
previous star formation. We derive an integrated oxygen abundance
of 12+log(O/H)=7.53$\pm$0.06 and a nitrogen-to-oxygen ratio of log(N/O)=-1.57$\pm$0.19.
Our results are consistent, within the uncertainties, with a homogeneous
distribution of oxygen and nitrogen within the ISM of the galaxy.
The high-ionization He\,{\sc ii} $\lambda$4686 emission line is detected only in the central part of  
HS\ 2236+1344. Similar to many BCDs with He\,{\sc ii} $\lambda$4686 emission, HS\ 2236+1344 shows no Wolf-Rayet (WR) 
bump.
}
   {}


    \titlerunning{GMOS-IFU spectroscopy of the XBCD galaxy HS\ 2236+1344}
    \authorrunning{Lagos et al.}            
   
   \maketitle
%
%________________________________________________________________

% ==============================================================================================
\section{Introduction}
% ==============================================================================================

Low-mass galaxies undergoing a violent burst of star formation, such as BCD galaxies 
(Thuan \& Martin \cite{ThuanMartin1981}, Loose \& Thuan \cite{LT86}) in the local Universe, and analogous objects 
at higher redshift (e.g., \emph{compact narrow emission-line galaxies} and \emph{green peas}, 
Guzm\'an et al. \cite{Guzman1998}, Cardamone et al. \cite{Cardamone2009}, Amor\'in et al. \cite{Amorin2012}) 
are important testbeds of galaxy evolution. 
In particular, the most metal-poor of these systems (e.g., Papaderos et al. \cite{P08} and references therein)
are of special importance, as they are the closest analogs of the low-mass galaxy building blocks that are
thought to have formed in the early Universe.

BCDs, the low-mass, high-compactness members of the broader class of 
H\,{\sc ii} galaxies (Terlevich et al. \cite{Terlevich1991}, Telles et al. \cite{T97}), span a wide range in
morphology and gas-phase metallicity (Kunth \& Sargent \cite{KS83}, Izotov \& Thuan \cite{IT99}). 
In their majority, these systems have subsolar oxygen abundance 
7.9$\la$12+log(O/H)$\la$8.4 (Terlevich et al. \cite{Terlevich1991}, Izotov \& Thuan \cite{IT99}, 
Kunth \& \"{O}stlin \cite{KO00}) and fall into the iE-- or nE type of the classification scheme by
Loose \& Thuan (\cite{LT86}), due to the presence of one or several luminous SF
regions in the central part of an old, more extended stellar low-surface
brightness (LSB) host (Papaderos et al. \cite{P96a}, Telles \& Terlevich \cite{TT97}, 
Cair\'os et al. \cite{C01}, Bergvall \& \"{O}stlin \cite{BO02},
Gil de Paz \cite{Gil03}, Amor\'in et al. \cite{Amorin09}, see Bergvall \cite{Bergvall12} for a review).
A smaller fraction ($\sim$10\%) of BCDs exhibit \emph{cometary} morphology (iI,C in the Loose \& Thuan (\cite{LT86})
classification) that is owing to strong ongoing SF activity at the one tip of an elongated stellar LSB `tail'. 

The lowest-metallicity BCDs (XBCDs), or extremely metal-poor (XMP) BCDs, defined as systems with an oxygen abundance 
12+log(O/H)$\la$7.6 (Izotov \& Thuan \cite{IT99}), are the least chemically evolved emission-line 
galaxies currently known. 
These systems are very scarce in the nearby Universe (Izotov \& Thuan \cite{IT07}, Kniazev et al. \cite{Kniazev04},
Papaderos et al. \cite{P08}, Guseva et al. \cite{G09}, see also Skillman \cite{Skillman12-JENAM})
with only about 100 of them currently known (see Morales-Luis et al. \cite{ML11}, Filho et al. \cite{Filho13} 
for recent compilations of literature data).
The most metal-poor ones are \object{SBS\ 0335-052\,W} (12+log(O/H)$\simeq$6.9\dots7.12, Izotov et al. \cite{I05},
Papaderos et al. \cite{P06a}, Izotov et al. \cite{I09}), 
\object{SBS\ 0335-052\,E}  (12+log(O/H)=7.2\dots7.3, Izotov et al. \cite{I97a}, Papaderos et al. \cite{P06a}) and 
\object{I\ Zw\ 18} (12+log(O/H)=7.2, Izotov et al. \cite{I97b}).
As pointed out by Papaderos et al. (\cite{P08}), BCDs/XBCDs show an intriguing connection 
between gas-phase metallicity, morphology and evolutionary status, with cometary morphology 
being a typical characteristic of the most metal-poor systems. 
This, together with the low ($\ga$1~Gyr) luminosity-weighted age of XBCDs suggests that most 
of their stellar mass has formed in the past 1--3 Gyr (Papaderos et al. \cite{P08}). 
Interestingly, a significant fraction ($\geq$10\%) of unevolved, high-redshift galaxies in the Hubble Deep Field 
also show cometary (also referred to as `tadpole') morphology 
(van den Bergh et al. \cite{v96}, Elmegreen et al. \cite{Elmegreen05}, Straughn et al. \cite{Straughn06},
Windhorst et al. \cite{Windhorst06}, Elmegreen et al. \cite{DME07}).

The formation process of cometary field galaxies near and far is unclear. Proposed hypotheses range from 
propagating SF, in the case of XBCDs (Papaderos et al. \cite{P98}), to interactions 
(Straughn et al. \cite{Straughn06}) and gas instabilities 
in forming disks (Elmegreen et al. \cite{Elmegreen05,ED12}) in the case of more massive galaxies.
In fact, all these mechanisms may be at work at some level. For example, whereas H\,{\sc ii}/BCDs are spatially well 
separated from normal (Hubble-type) galaxies (Telles \& Maddox \cite{T00}), they usually have low-mass 
stellar and/or gaseous companions (e.g., Taylor et al. \cite{Taylor95}, Noeske et al. \cite{Noeske01}).
This also applies to many XBCDs, many of which are known to reside within loose galaxy groups 
(Pustilnik et al. \cite{Pustilnik01}) or in galaxy pairs with a linear separation between 
$\sim$2 kpc and $\sim$100 kpc (Papaderos \cite{P12}, hereafter P12). 
Examples of such binary dwarf galaxies are, e.g. I\ Zw\ 18 (Lequeux \& Viallefond \cite{LV80}) and 
SBS 0335-052 (Papaderos et al. \cite{P98}, Pustilnik et al. \cite{Pustilnik01-SBS0335}).
Recently, P12 pointed out the intriguing similarity of some XBCDs with their 
low-mass companions in their structural properties and evolutionary status, and interpreted this  
as manifestation of synchronization in the assembly history of such binary galaxies. 
He proposed, based on heuristic arguments, that the delayed formation of some present-day XBCDs in pairs or groups 
can qualitatively be explained as the result of their co-evolution within a common dark matter (DM) halo: 
The cumulative effect of recurrent mild interactions between co-evolving dwarfs is, according to P12, a 
quasi-continuous heating of their gas component and the delay of the dominant phase of their formation, 
in agreement with the observed low chemical abundances and young luminosity-weighted ages of XBCDs.

On sub-galactic scales, observations of H\,{\sc ii}/BCDs also hint at a synchronous 
star formation history (SFH). For example, near-IR photometry and optical spectroscopy studies of 
\object{UM~461} and \object{Mrk~36} by Lagos et al. (\cite{L11}) and \object{UM 408} 
by Lagos et al. (\cite{L09,L11}) reveal that Young Stellar Clusters (YSCs) 
in these galaxies have a similar age over a spatial scale of $\sim$1 kpc (see, e.g., Telles \cite{Telles10}). 
Evidence for coeval star formation on scales of several 
hundred pc has also been found through photometric studies of high-luminosity BCDs with the Hubble Space Telescope 
(e.g., \"{O}stlin et al. \cite{Ostlin03}, Adamo et al. \cite{Adamo11}) and surface photometry studies.
For example, the subtraction of the underlying LSB host emission from surface brightness profiles 
of the BCD \object{Mrk~178} yields almost flat color gradients within the starburst component, 
in agreement with the picture of a nearly coevally triggered SF episode 
on spatial scales of $\sim$1 kpc (Papaderos et al. \cite{Papaderos02}). 

With regard to the chemical abundance patterns of H\,{\sc ii}/BCDs, the available 
data suggest that the metals from previous SF events are homogeneously
distributed and mixed over the ISM, while freshly produced elements 
by massive stars remain unmixed with the warm ISM and reside in the hot gas 
phase (Tenorio-Tagle \cite{TT96}, Kobulnicky \& Skillman \cite{KS97}).
The spatial constancy of the N/O ratio, observed in these galaxies, 
might also be attributed to efficient transport and mixing of metals 
(e.g., P\'erez-Montero et al. \cite{PM11}, Lagos et al. \cite{L09,L12}) by hydrodynamical processes 
(e.g., starburst-driven flows, infall of gas), so keeping the N/O ratio constant through the ISM 
of the galaxies on large scales.
Locally enhanced N/O ratios in some BCDs are probably explained by nitrogen
enrichment through WR winds (Walsh \& Roy \cite{WR89}, Kobulnicky et al. \cite{K97}), resulting in an N overabundance
at the location of YSCs, in agreement with the detection of small, yet statistically 
significant oxygen abundance variations, attributable to chemical self-enrichment (Kunth \& Sargent \cite{KS86})
even in some XBCDs (Papaderos et al. \cite{P06a}).

% ===================== Figure 1 ====================================================================
\begin{figure}
\centering
\includegraphics[width=9cm]{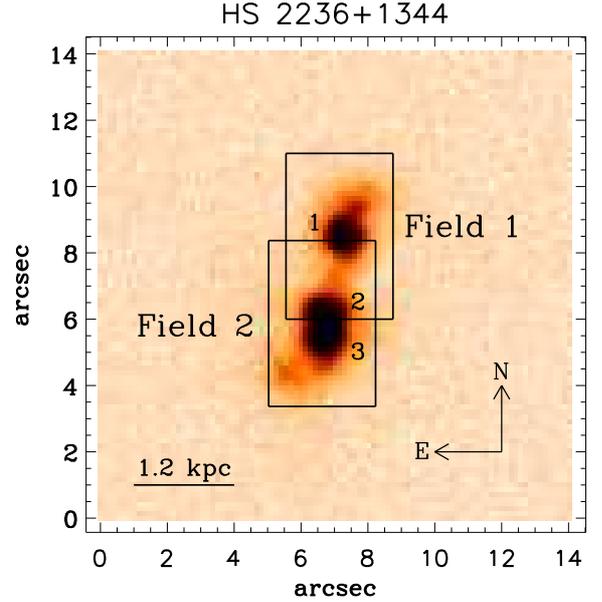}
\caption{g-band acquisition image of the galaxy HS\ 2236+1344.  
The rectangles indicate the position of the two GMOS--IFU FoV of 3\farcs5$\times$5\arcsec.
The three GH\,{\sc ii}Rs, resolved in this study, are indicated in the figure.
}
\label{image_campo}
\end{figure}

Integral field unit (IFU) spectroscopy offers a powerful tool to study all above 
topical issues in BCD/XBCD research in spatial detail (an overview of all  H\,{\sc ii}/BCD galaxies studied thus far 
with IFU spectroscopy is shown in Lagos \& Papaderos \cite{LP13}). 
Our main objective in this paper is to carry out a spatially resolved investigation of the 
chemical abundance patterns and kinematics of the warm ISM in the nearby ($\sim$80 Mpc) 
XBCD galaxy HS\ 2236+1344. For this, we use medium-resolution IFU data obtained 
with the Gemini North telescope. 
In Fig.~\ref{image_campo} we show the $g$-band acquisition image of the galaxy
(see Table~\ref{parameters} for a summary of its general properties).
It can be seen that HS\ 2236+1344 contains three compact (regions 1, 2 and 3) high-surface brightness knots
or GH\,{\sc ii}Rs, and some faint arm-like features departing from its central body. 
These may be attributed to starburst-driven shells 
and/or tidally induced features, perhaps connected with an ongoing merger.

% ===================== Table 1 ====================================================================
\begin{table}
 \centering
   \caption{General parameters of HS\ 2236+1344.}
  \begin{tabular}{@{}lcc@{}}
  \hline   
parameter              &  value                    & reference \\
\hline
$\alpha$ (J2000)       & 22$^{h}$38$^{m}$31.1$^{s}$&(a)\\
$\delta$ (J2000)       & +14$^{o}$00$^{m}$30$^{s}$ &(a)\\
D (3K CMB)             & 79.7 Mpc                  &(a)\\ 
1\arcsec (3K CMB)      & 386 pc                    &(a)\\
z                      & 0.020628                  &(b)\\
12+log(O/H)            & 7.53                      &(b)\\
Log(M(H\,{\sc i})) M$_{\odot}$ &  $<$8.35                  &(c)\\
Other name             & SDSS J223831.12+140029.7  &\\       
\hline
\end{tabular}
\tablefoot{
\tablefoottext{a}{Obtained from NED,}
\tablefoottext{b}{Derived from the present observations,}
\tablefoottext{c}{Pustilnik \& Martin (\cite{PS07}).}
\label{parameters}
}
\end{table}

This paper is organized as follows: the observations and data reduction
are discussed in Sect.~\ref{observation}. In Sect.~\ref{results} we presented and discuss the results. 
Our conclusions are summarized in Sect.~\ref{summary}.

% ==========================================================
\section{Observations and Data reduction}\label{observation}
% ==========================================================

The observations discussed in this paper were performed using the Gemini
GMOS (Hook et al. \cite{Hook04}) 
and the IFU unit, hereafter GMOS--IFU (Allington et al. \cite{Allington02}),
at Gemini North telescope, using the grating R600$+\_$G5304 (R600) in one-slit mode 
with a spectral resolution of R$\sim$4000. 
The GMOS-IFU in one slit mode composes a pattern
of 500 hexagonal elements, each with a projected diameter of 0\farcs2, 
covering a total field of view (FoV) of 3\farcs5$\times$5$\arcsec$, plus 250 elements sampling the sky. 
The data were obtained at low airmass during the same night but in two
different positions across the galaxy as indicated in Fig.~\ref{image_campo} 
(cf Table~\ref{observation}).

% ===================== Table 2 ====================================================================
\begin{table}
 \centering
  \caption{Observing log.}
  \begin{tabular}{@{}lccccclrlr@{}}
  \hline      
 Name   & Date &Exp. time & Airmass\footnote{1}\\
        &      & (s)      &         \\
 \hline
Field 1     & 2010-09-08&3$\times$1285& 1.068\\ 
Field 2     & 2010-09-08&3$\times$1285& 1.013\\
\hline
\end{tabular}
\tablefoot{
\tablefoottext{1}{Obtained from the average values of the different exposures.}
}
\end{table}

The data reduction was carried out using the Gemini software package version 1.9 within  
IRAF\footnote{IRAF is distributed  by NOAO, which is operated by the Association of
Universities for Research in Astronomy Inc., under cooperative agreement
with the National Science Foundation.}. This includes bias subtraction, 
flat-field correction, wavelength calibration and sky subtraction. 
The flux calibration was performed using the sensitivity function derived from
observation of the spectrophotometric standard star BD+28d4211.
The 2D data images were transformed into 3D data cubes, resampled to a 
0\farcs2 spatial resolution, and corrected for differential atmospheric refraction
using the \textit{gfcube} routine. Finally, the two data cubes obtained with the grating
R600 were combined into a final data cube that covers a spectral 
range from $\sim$4300 $\rm\AA$ to 6800 $\rm\AA$ and a FoV of 4\arcsec$\times$8\arcsec (20$\times$40 spaxels).
More details about the data reduction can be found in Lagos et al. (\cite{L09,L12}).

% ===================== Figure 2 ====================================================================
\begin{figure*}
\centering
\includegraphics[width=90mm]{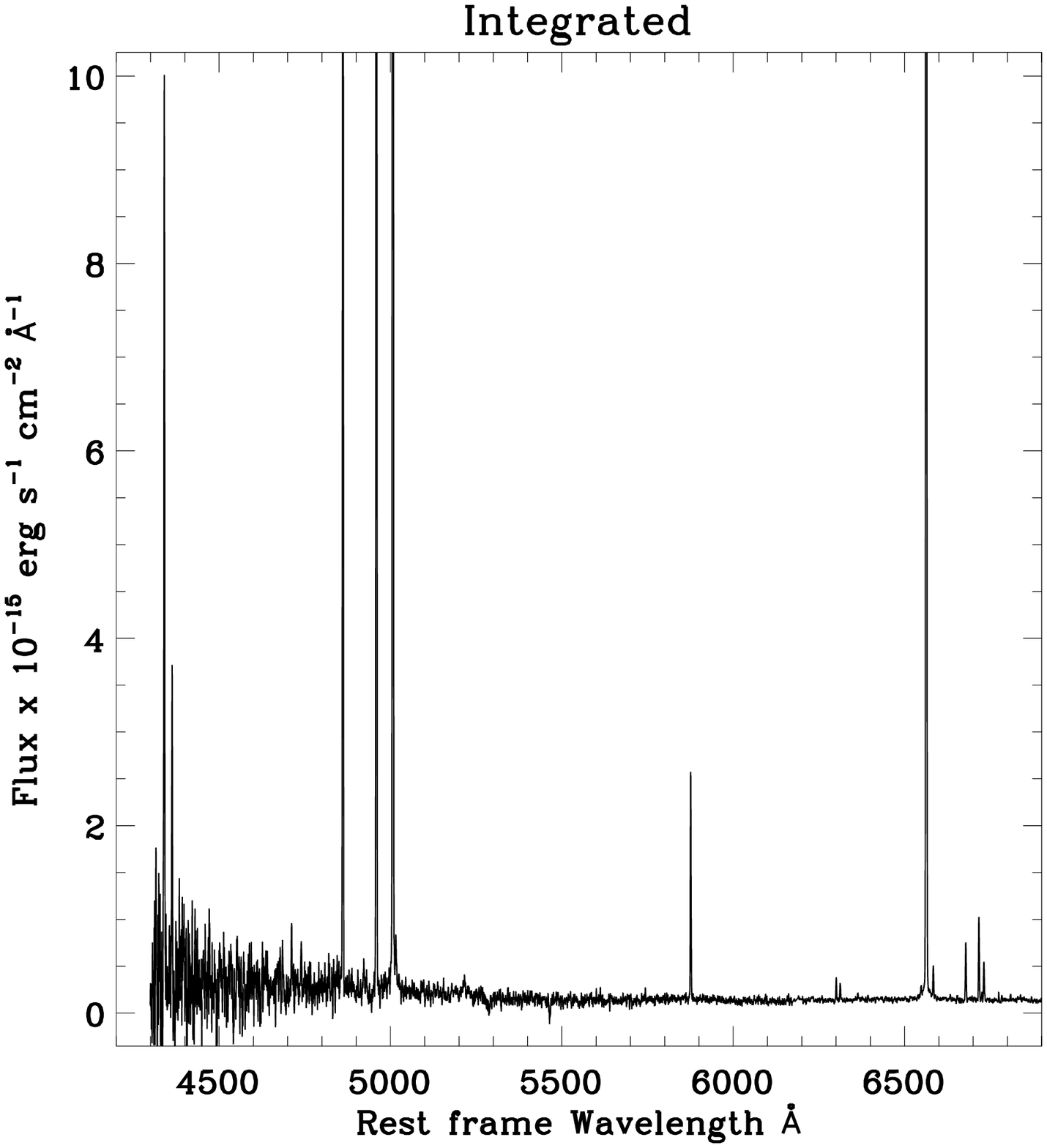}
\includegraphics[width=90mm]{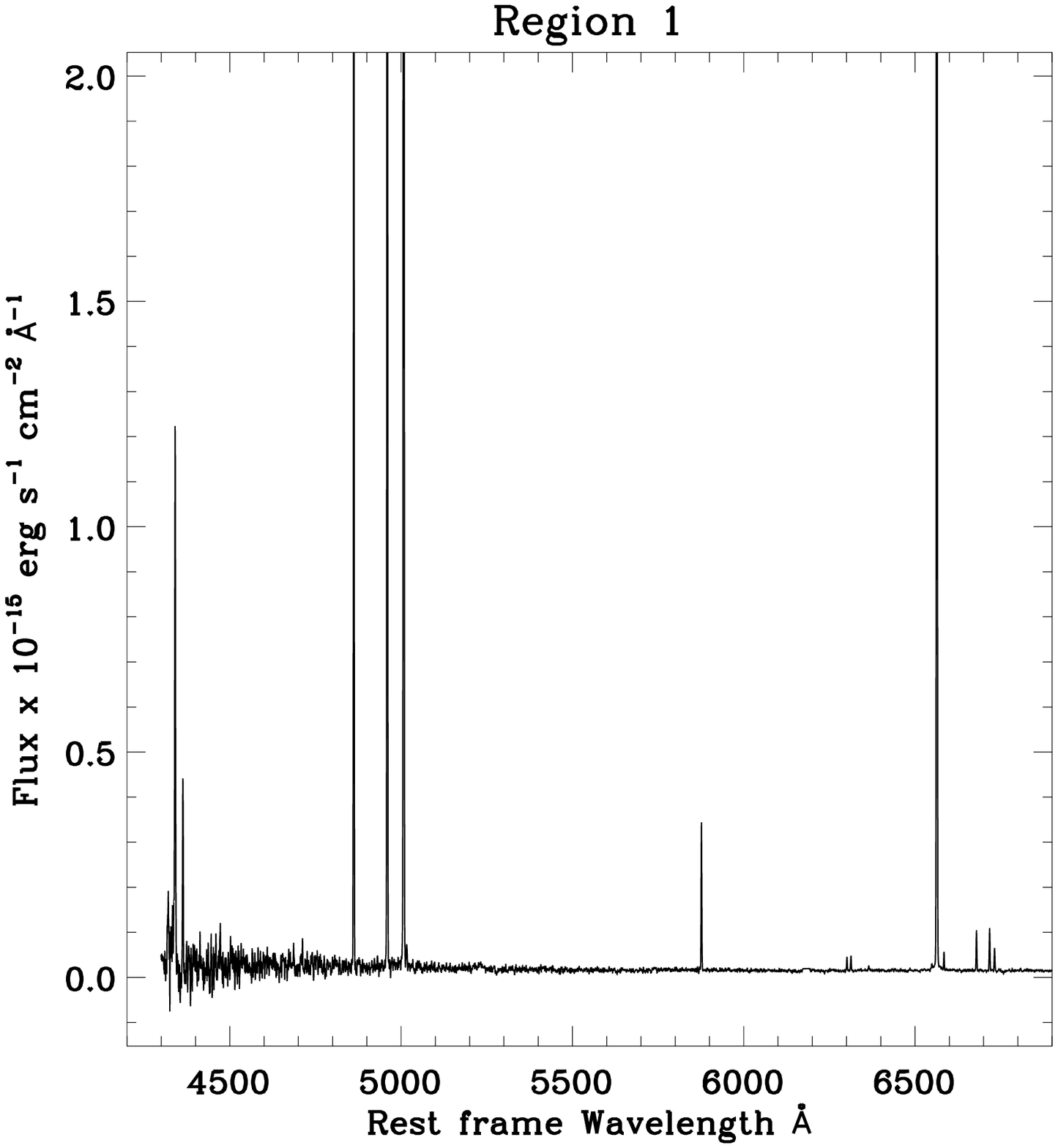}\\
\includegraphics[width=90mm]{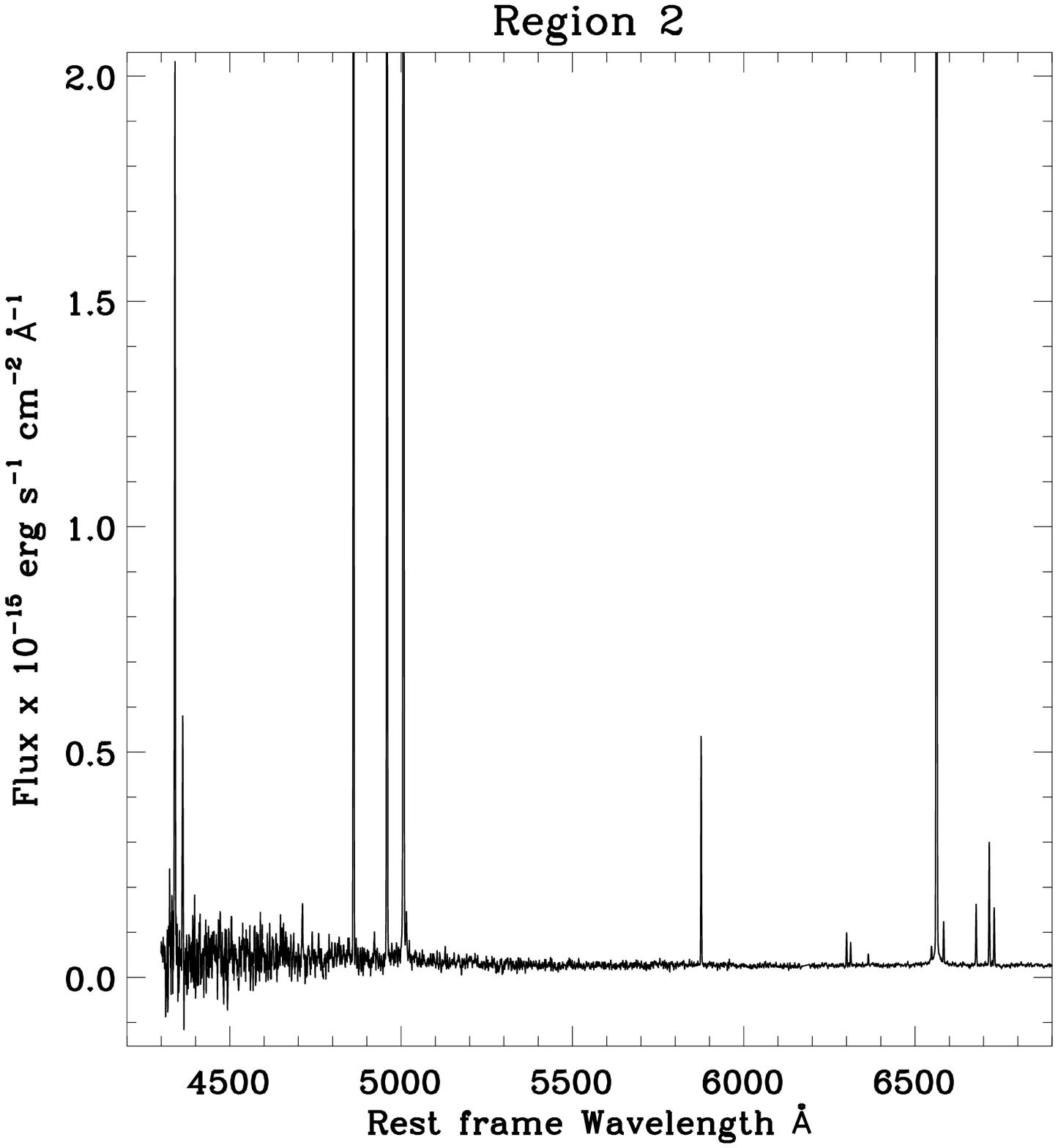}
\includegraphics[width=90mm]{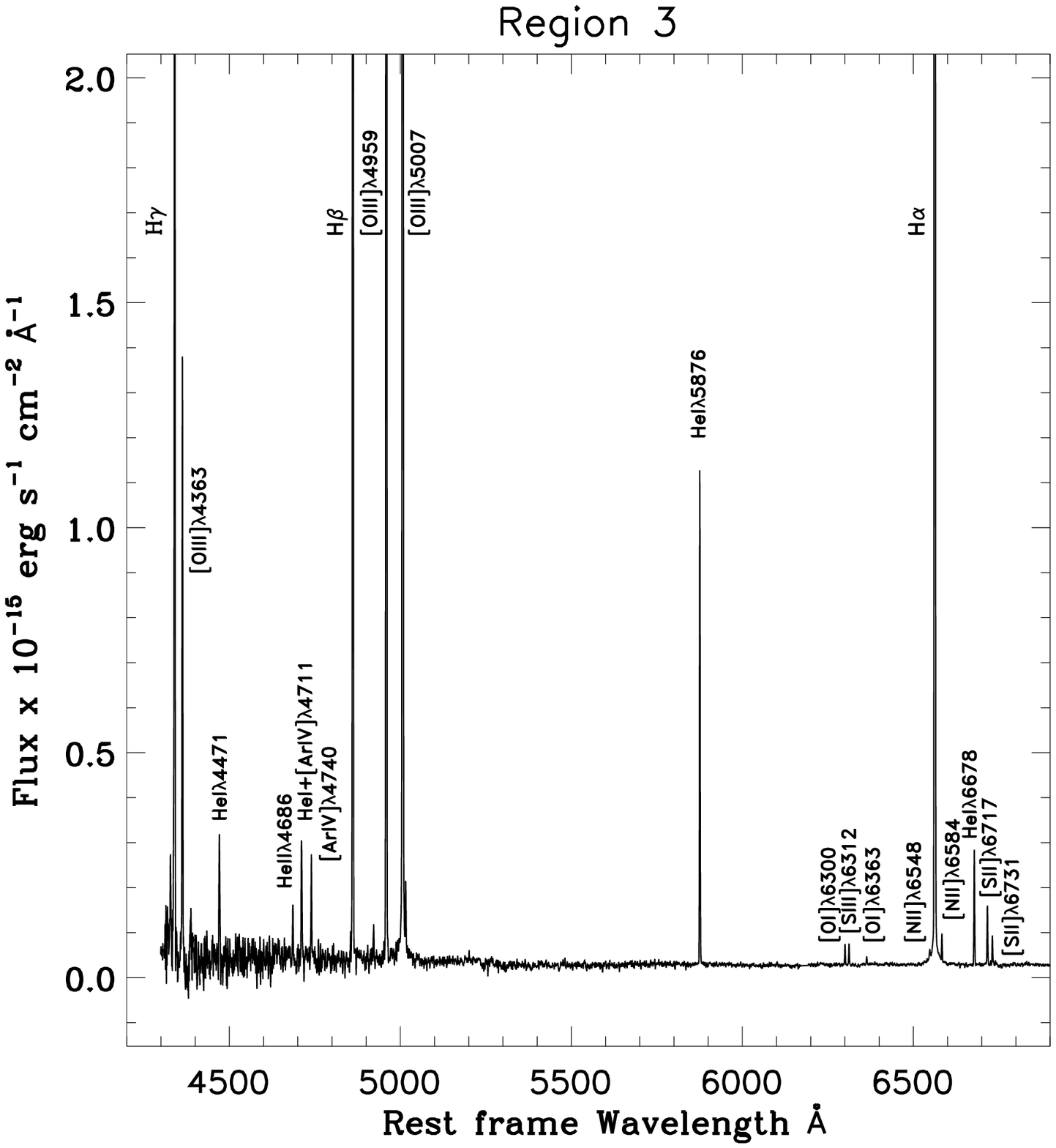}

\caption{Integrated Gemini GMOS spectrum of HS\ 2236+1344, and spectra extracted from regions 1--3.
Note the presence of a broad low-intensity emission component in H$\alpha$
(possibly also in H$\beta$) and in the forbidden [O\,{\sc iii}]$\lambda$5007 line.}
\label{integrated_spec}
\end{figure*}

Figure \ref{integrated_spec} shows the integrated spectrum of the galaxy (over the whole FoV) 
and the spectra of regions 1, 2 and 3, summed up over all spaxels within the areas indicated 
in Sect.~\ref{emission}.
In this figure we label the strongest emission lines detected and used in our study.
The most remarkable finding from our data is the localized He\,{\sc ii}
$\lambda$4686 emission in region 3 only (also visible on the integrated spectrum of HS\ 2236+1344) 
and a broad component in the base of some emission lines, such as H$\alpha$ and [O\,{\sc iii}] $\lambda$5007. 
Finally, the emission line fluxes were measured using the IRAF task \textit{fitprofs} 
by fitting Gaussian profiles. Since most of the emission lines were measured 
using an automatic procedure, we assigned the value 0.0
to all spaxels with signal to noise ratio (S/N) $<$ 3.

% ==============================================================================================
\section{Results and discussion}\label{results}
% ==============================================================================================

\subsection{Emission lines, continuum and extinction structure} \label{emission}

\begin{figure*}
%\centering
\includegraphics[width=61mm]{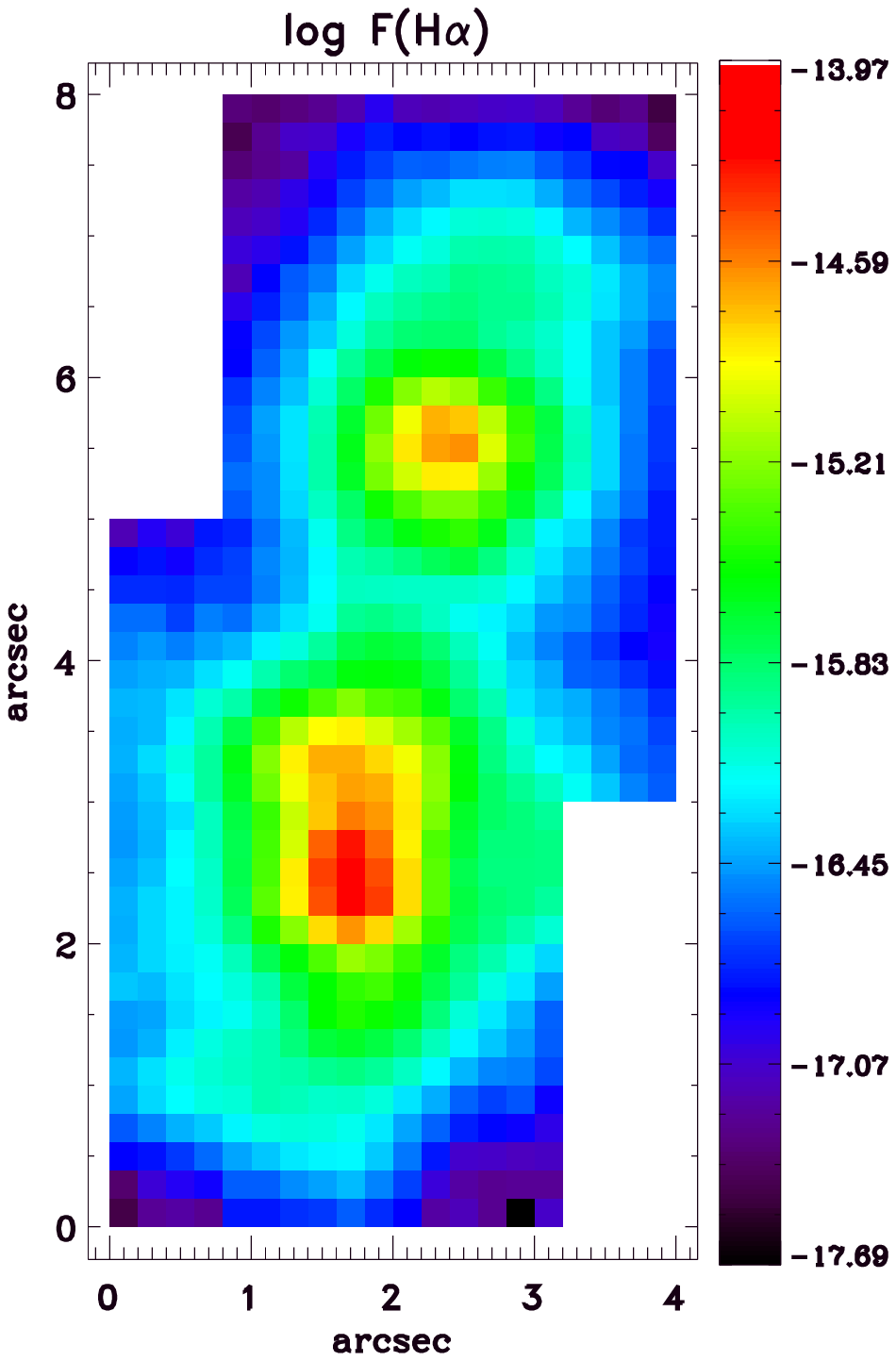}
\includegraphics[width=61mm]{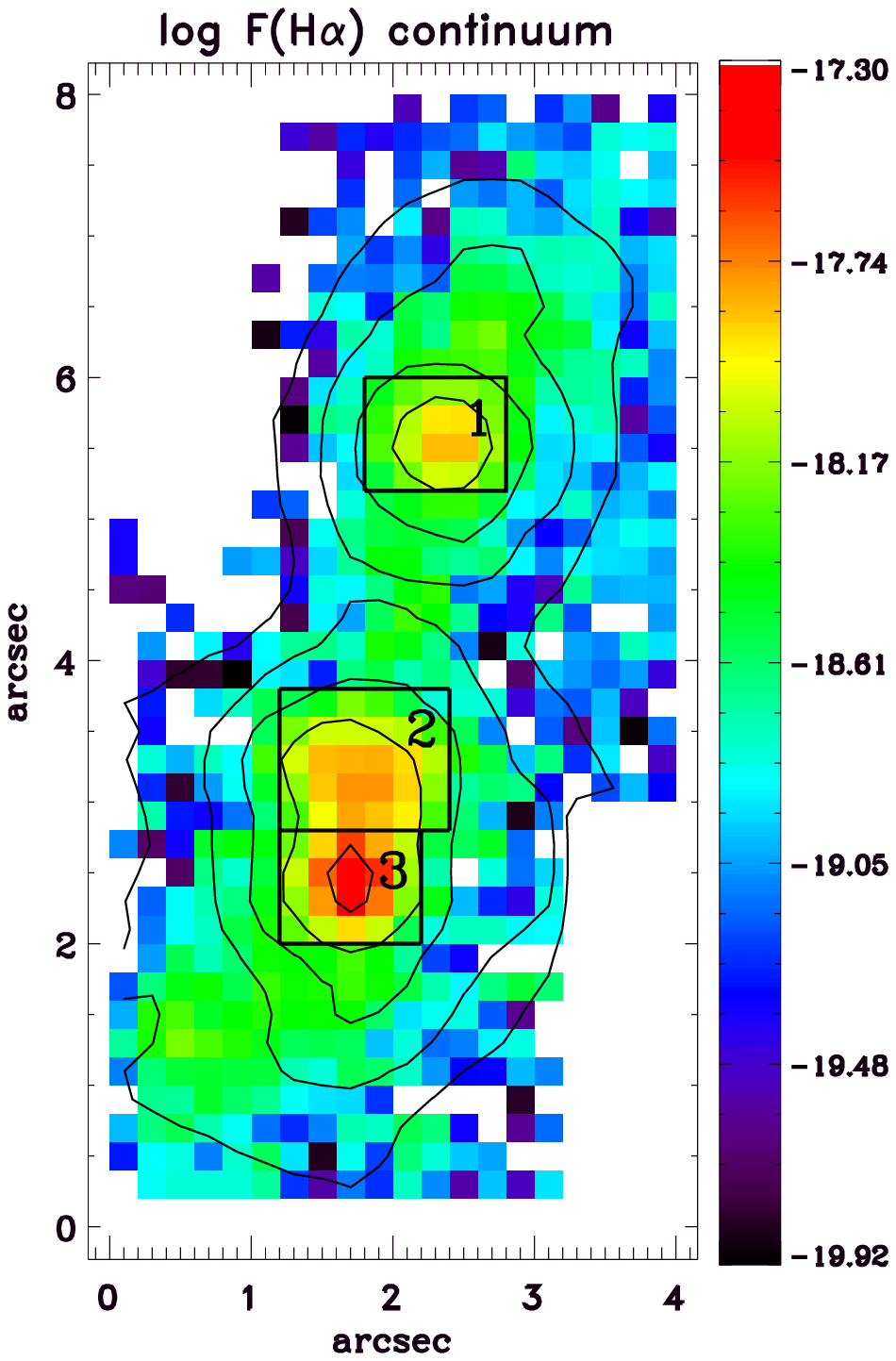}
\includegraphics[width=61mm]{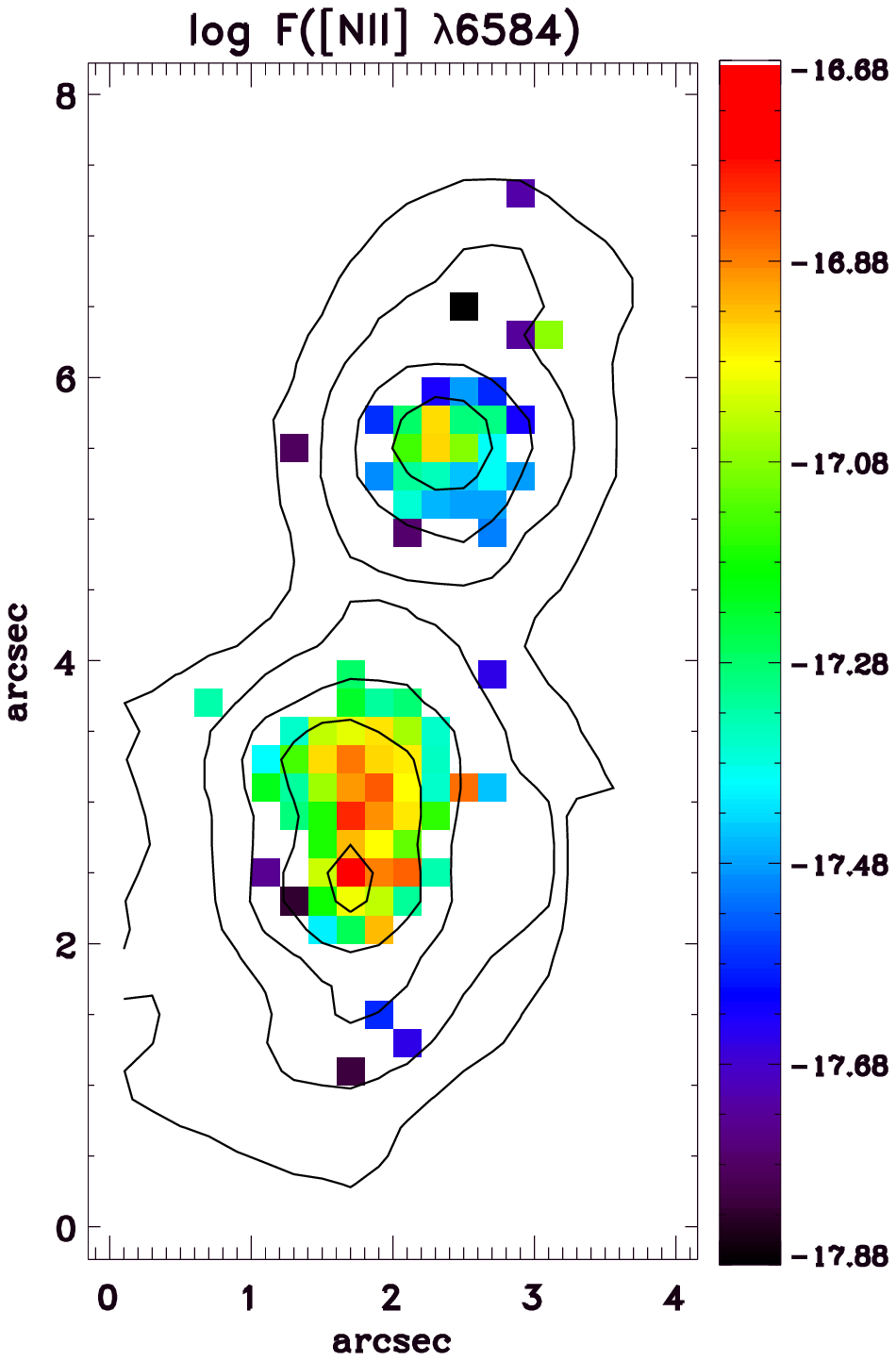}\\
\includegraphics[width=61mm]{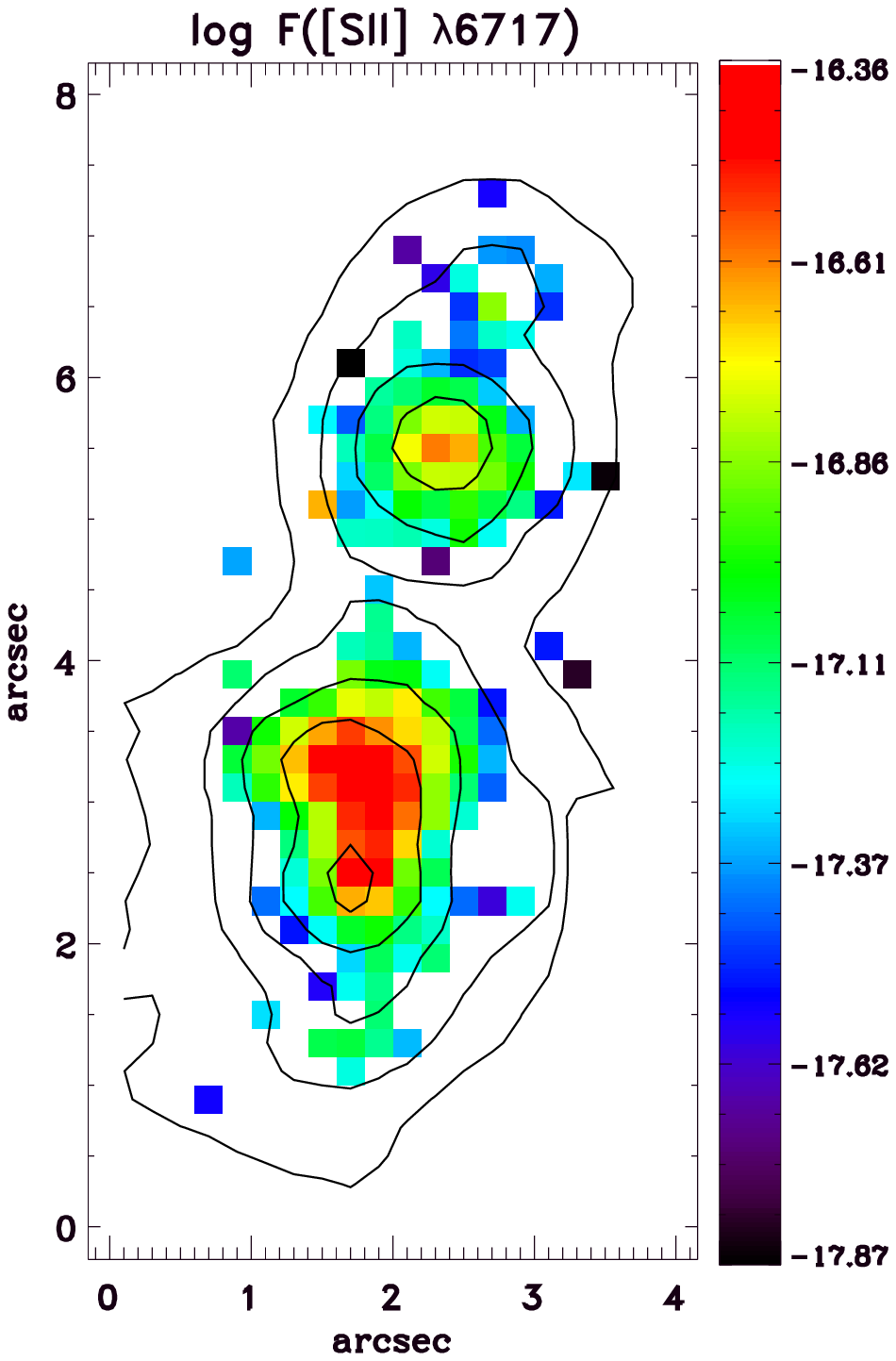}
\includegraphics[width=61mm]{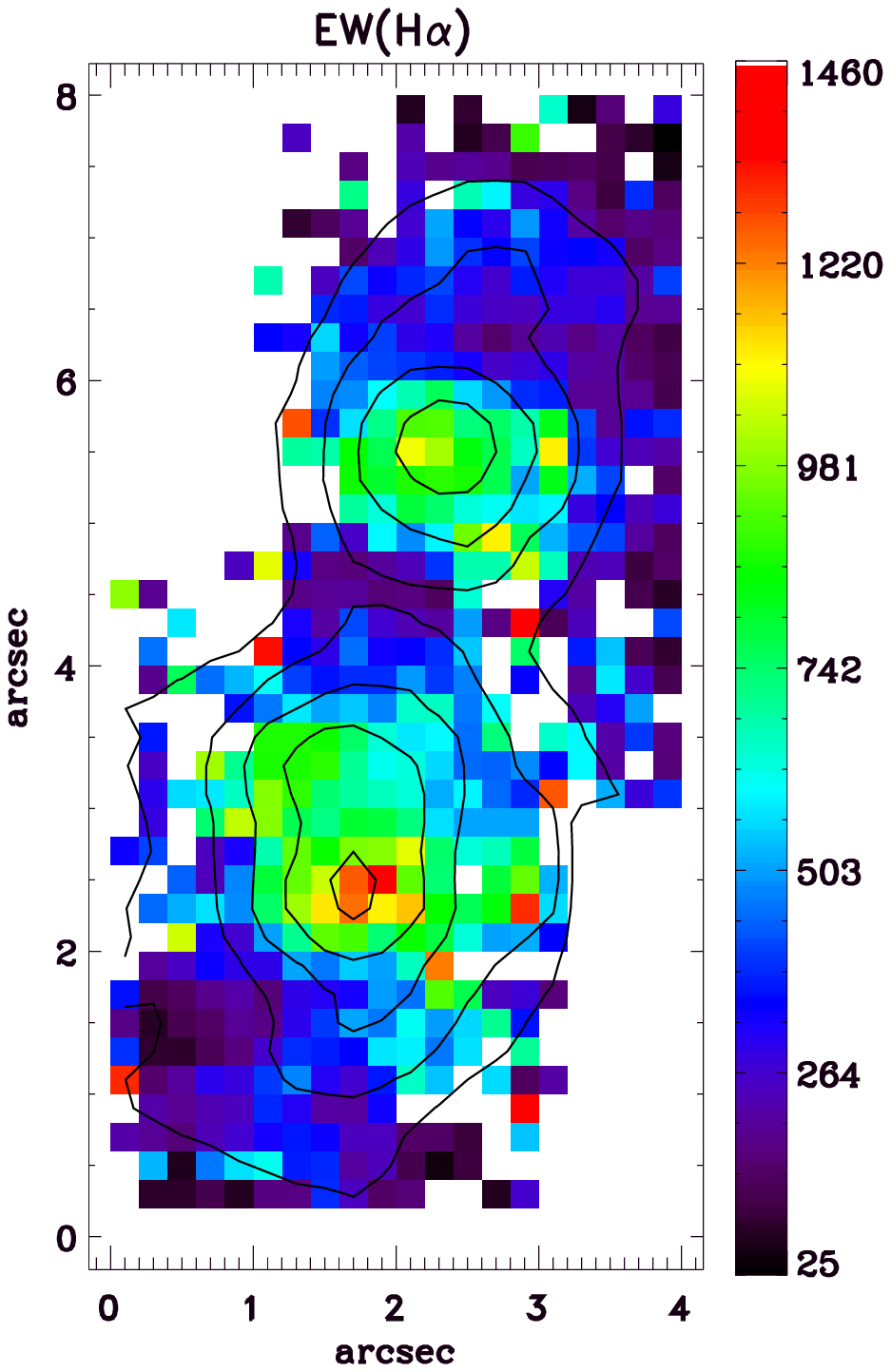}
\includegraphics[width=61mm]{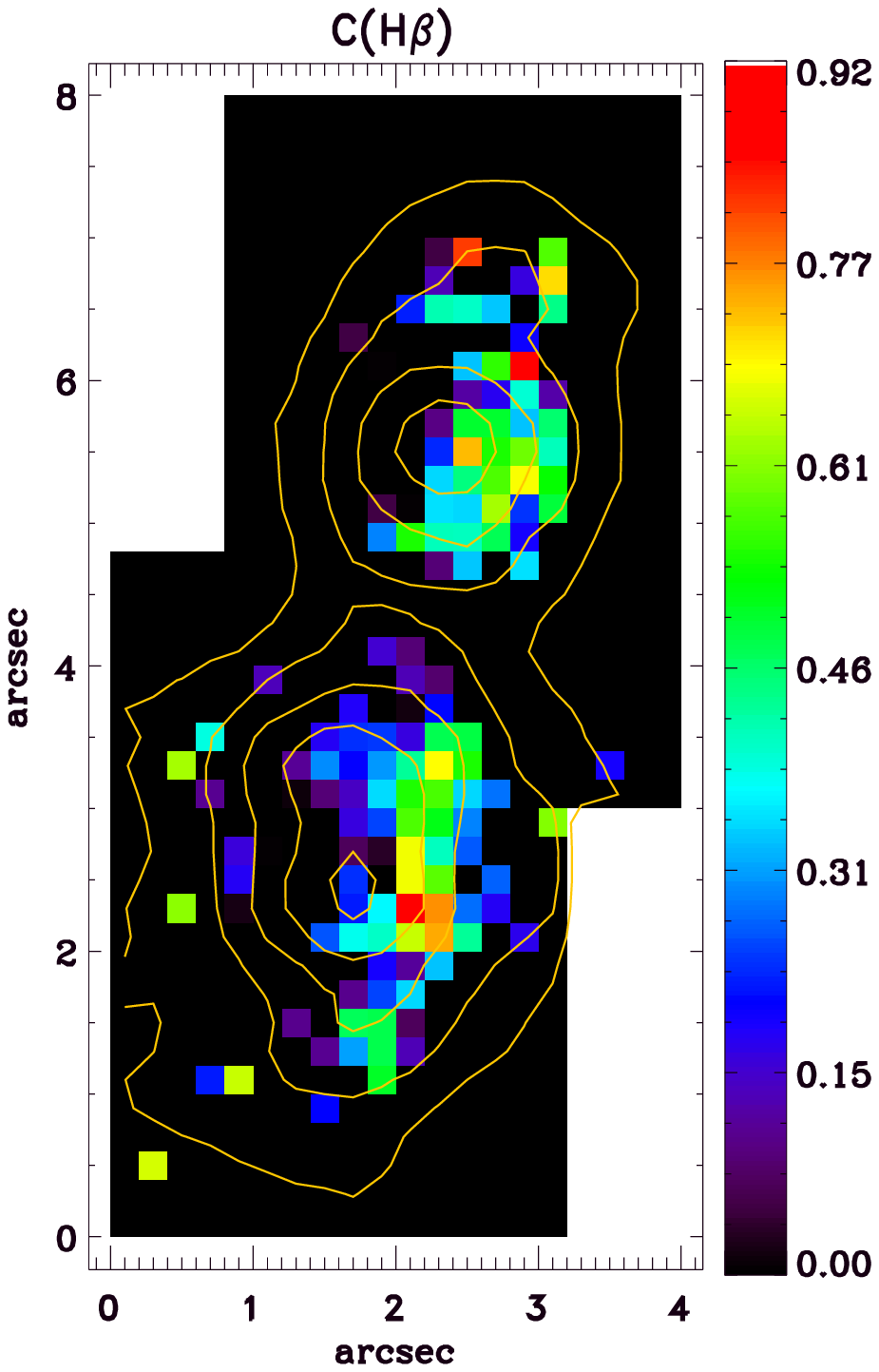}
\caption{
Logarithmic emission line map of the H$\alpha$ (top-left), H$\alpha$ continuum (top-middle), 
[N\,{\sc ii}] $\lambda$6584 (top-right), [S\,{\sc ii}] $\lambda$6717 (bottom-left), 
EW(H$\alpha$) (bottom-middle) and extinction c(H$\beta$) (bottom-right). 
Logarithmic emission-line fluxes in units of erg cm$^{-2}$ s$^{-1}$, the logarithmic continuum 
flux density is in erg cm$^{-2}$ s$^{-1}$ $\rm \AA^{-1}$
and equivalent widths in $\rm \AA$. Overlaid are the H$\alpha$ flux contours. 
In the top-middle (H$\alpha$ continuum) panel we indicate the different apertures
used in this study for regions 1, 2 and 3, respectively.
}
\label{mapas}
\end{figure*}

We used the flux measurement procedure described previously in Sect.~\ref{observation}
to construct emission line maps (e.g., [S\,{\sc ii}] $\lambda$6731, [S\,{\sc ii}] $\lambda$6717, 
[N\,{\sc ii}] $\lambda$6584, H$\alpha$, [O\,{\sc iii}] $\lambda$5007, 
[O\,{\sc iii}] $\lambda$4959, H$\beta$, [O\,{\sc iii}] $\lambda$4363, 
H$\gamma$) within the combined FoV of 4\arcsec$\times$8\arcsec. 
In Fig.~\ref{mapas} we show the H$\alpha$, H$\alpha$ continuum, [N\,{\sc ii}] $\lambda$6584, 
[S\,{\sc ii}] $\lambda$6717, EW(H$\alpha$) and the extinction c(H$\beta$) maps of the galaxy. 
The H$\alpha$ continuum map shows that HS\ 2236+1344 is composed of three GH\,{\sc ii}Rs. 
We designate the GH\,{\sc ii}R that is centered on the field~1 (see Fig.~\ref{image_campo}) as region 1, 
while the GH\,{\sc ii}R in the field 2 is resolved into a double peak that we labeled as
regions 2 and 3. These three GH\,{\sc ii}Rs are also resolved on the acquisition image
of the galaxy (Fig.~\ref{image_campo}).
We see that the structure of the ionized gas, as traced by the emission
lines and the H$\alpha$ and H$\beta$ equivalent widths are similar, without any offset 
compared with the distribution of the stellar continuum (see Figure \ref{mapas}). 

The extinction was calculated from the emission line ratio F(H$\alpha$)/F(H$\beta$)=2.76,
assuming case~B recombination (Osterbrock \& Ferland \cite{O06}) at T = 2 $\times$ 10$^4$ K
(see Table \ref{integrated_properties_1}).
The c(H$\beta$) values in Fig. \ref{mapas} ranges between 0 and $\sim$0.92 and shows 
an inhomogeneous pattern, with its peak value being slightly offset
from that of the H$\alpha$ emission.
Most of the spaxels in the outer part of the galaxy show a F(H$\alpha$)/F(H$\beta$) 
ratio that is close to the theoretical value, 
so we assumed that these regions do not suffer from intrinsic extinction.
Our results support the conclusion by Cair\'os et al. (\cite{C09}) and Lagos et al. (\cite{L09}) that 
a constant correction term for the extinction over the whole galaxy
may in some cases lead to large errors in the derivation of properties 
of individual regions in H{\sc ii} and BCD galaxies.

% ===================== Figure 3 ====================================================================

Finally, the observed emission line fluxes, for all the spaxels in our FoV, 
have been corrected for extinction using the observed Balmer
decrement and the Cardelli et al. (\cite{C98}) reddening curve f($\lambda$) 
as $I(\lambda)/I(H\beta)=F(\lambda)/F(H\beta) \times 10^{c(H\beta)f(\lambda)}$.
The observed F($\lambda$) and corrected emission line fluxes I($\lambda$) 
relative to the H$\beta$ and their errors, multiplied by a factor of 100, 
the observed flux of the H$\beta$ emission line, 
the EW(H$\alpha$) and EW(H$\beta$), and the extinction coefficient c(H$\beta$) 
for the integrated galaxy are tabulated in Table \ref{tb_lines_integrated}. 
In Table \ref{tb_lines} we list the same quantities for the three SF regions in the BCD.
We note that the mean extinction c(H$\beta$)$\simeq$0.20 we derive over 
HS\ 2236+1344 is in good agreement with the value of 0.157 derived by Thuan \& Izotov (\cite{TI05}). 
Determinations for individual regions differ, however, from the values of 0.49 and 0.27 obtained 
by Izotov \& Thuan (\cite{IT07}) from the H$\gamma$/H$\beta$ ratio for their regions 1 and 2, respectively.   

% ===================== Table 3 ====================================================================
\begin{table}
 \centering
 \begin{minipage}{80mm}
  \caption{Observed and extinction corrected emission line fluxes
  for the integrated spectrum of the galaxy. The fluxes are relative to F(H$\beta$)=100.}
  \begin{tabular}{@{}lcccccccc@{}}
  \hline
       &   \multicolumn{2}{c}{Integrated}  \\
       & F($\lambda$)/F(H$\beta$)  & I($\lambda$)/I(H$\beta$) \\
 \hline
H$\gamma  \lambda$4340            &  46.12$\pm$3.88 &  49.21$\pm$7.31\\
$\left[OIII\right] \lambda$4363   &  14.75$\pm$0.95 &  15.69$\pm$1.36\\
HeI $\lambda$4471                 &$>$4.30$\pm$0.39 &$>$4.51$\pm$0.59\\
HeII $\lambda$4686                &$>$1.20$\pm$0.20 &$>$1.22$\pm$0.29\\
HeI+$\left[ArV\right] \lambda$4711&$>$3.80$\pm$0.53 &$>$3.87$\pm$0.76\\
$\left[ArV\right] \lambda$4740    &$>$2.70$\pm$0.14 &$>$2.74$\pm$0.20\\
H$\beta \lambda$4861              & 100.00$\pm$2.71 & 100.00$\pm$3.83\\
$\left[OIII\right] \lambda$4959   & 169.96$\pm$5.02 & 168.61$\pm$7.05\\
$\left[OIII\right] \lambda$5007   & 507.75$\pm$12.70& 499.87$\pm$17.68\\
HeI $\lambda$5876                 &  11.92$\pm$0.68 &  10.96$\pm$0.88\\
$\left[OI\right] \lambda$6300     &   1.03$\pm$0.20 &   0.92$\pm$0.25\\
$\left[SIII\right] \lambda$6312   &   1.10$\pm$0.12 &   0.99$\pm$0.14\\
$\left[OI\right] \lambda$6363     &   0.32$\pm$0.03 &   0.29$\pm$0.04\\
$\left[NII\right] \lambda$6548    &   0.61$\pm$0.10 &   0.54$\pm$0.16\\
H$\alpha \lambda$6563             & 317.83$\pm$6.25 & 280.95$\pm$7.81\\
$\left[NII\right] \lambda$6584    &   2.05$\pm$0.14 &   1.81$\pm$0.17\\
HeI $\lambda$6678                 &   3.43$\pm$0.45 &   3.01$\pm$0.56\\
$\left[SII\right] \lambda$6717    &   4.94$\pm$0.30 &   4.33$\pm$0.37\\
$\left[SII\right] \lambda$6731    &   2.13$\pm$0.38 &   1.86$\pm$0.47\\
                                  &                                  \\
F(H$\beta$)\footnote{In units of $\times$10$^{-14}$ erg cm$^{-2}$ s$^{-1}$} &5.16$\pm$0.07 \\
EW(H$\alpha$)\footnote{In units of $\rm \AA$}                               &834   \\ 
EW(H$\beta$)$^b$                                                            &157   \\ 
c(H$\beta$)                                                                 &0.18 \\
\hline
\end{tabular}
\label{tb_lines_integrated}
\end{minipage}
\end{table}

% ===================== Table 4 ====================================================================
\begin{table*}
 \centering
 \begin{minipage}{180mm}
  \caption{Observed and extinction corrected emission lines for the regions 1, 2 and 3. 
The fluxes are relative to F(H$\beta$)=100.}
  \begin{tabular}{@{}lcccccccc@{}}
  \hline
       &   \multicolumn{2}{c}{Region 1} &   \multicolumn{2}{c}{Region 2}&   \multicolumn{2}{c}{Region 3}\\
       & F($\lambda$)/F(H$\beta$)  & I($\lambda$)/I(H$\beta$) & F($\lambda$)/F(H$\beta$) & I($\lambda$)/I(H$\beta$)& F($\lambda$)/F(H$\beta$) & I($\lambda$)/I(H$\beta$)\\
 \hline
H$\gamma\  \lambda$4340           & 45.15$\pm$2.27 &48.70$\pm$3.46  & 46.70$\pm$2.93 & 50.01$\pm$4.44 & 47.80$\pm$2.89 & 49.56$\pm$4.24\\
$\left[OIII\right] \lambda$4363   & 16.29$\pm$0.86 &17.51$\pm$1.31  & 13.76$\pm$0.89 &14.69$\pm$1.34  & 17.43$\pm$0.64 & 18.04$\pm$0.94\\
HeI $\lambda$4471                 & $\cdots$       & $\cdots$       & $\cdots$       & $\cdots$       &  4.30$\pm$0.39 &  4.42$\pm$0.57\\
HeII $\lambda$4686                & $\cdots$       & $\cdots$       & $\cdots$       & $\cdots$       &  1.20$\pm$0.20 &  1.22$\pm$0.29\\
HeI+$\left[ArV\right] \lambda$4711& $\cdots$       & $\cdots$       & $\cdots$       & $\cdots$       &  3.80$\pm$0.53 &  3.84$\pm$0.76\\
$\left[ArV\right] \lambda$4740    & $\cdots$       & $\cdots$       & $\cdots$       & $\cdots$       &  2.70$\pm$0.14 &  2.72$\pm$0.20\\
H$\beta\ \lambda$4861             &100.00$\pm$1.22 &100.00$\pm$1.73 &100.00$\pm$3.69 &100.00$\pm$5.22 &100.00$\pm$3.14 &100.00$\pm$4.44\\
$\left[OIII\right] \lambda$4959   &179.60$\pm$7.18 &177.38$\pm$10.03&143.12$\pm$3.54 &141.52$\pm$4.95 &186.38$\pm$6.07 &185.28$\pm$8.53\\
$\left[OIII\right] \lambda$5007   &547.94$\pm$9.42 &538.03$\pm$13.08&431.19$\pm$9.75 &424.13$\pm$13.56&554.97$\pm$19.19&550.17$\pm$26.90\\
HeI $\lambda$5876                 &116.59$\pm$1.32 &105.68$\pm$1.69 & 10.92$\pm$0.38 &9.99$\pm$0.49   & 12.98$\pm$0.67 & 12.39$\pm$0.90\\
$\left[OI\right] \lambda$6300     &  1.10$\pm$0.13 &  0.97$\pm$0.16 &  1.58$\pm$0.12 &1.41$\pm$0.15   &  0.53$\pm$0.02 &  0.50$\pm$0.03\\
$\left[SIII\right] \lambda$6312   &  1.22$\pm$0.16 &  1.07$\pm$0.20 &  1.05$\pm$0.12 &0.93$\pm$0.15   &  0.54$\pm$0.02 &  0.51$\pm$0.03\\
$\left[OI\right] \lambda$6363     &  0.41$\pm$0.06 &  0.36$\pm$0.07 &  0.52$\pm$0.05 &0.46$\pm$0.06   &  0.17$\pm$0.03 &  0.16$\pm$0.04\\
$\left[NII\right] \lambda$6548    &  0.40$\pm$0.10 &  0.35$\pm$0.14 &  0.71$\pm$0.09 &0.62$\pm$0.19   &  0.26$\pm$0.06 &  0.24$\pm$0.08\\
H$\alpha\ \lambda$6563            &324.20$\pm$3.49 &280.75$\pm$4.27 &321.10$\pm$6.81 &281.90$\pm$8.45 &298.95$\pm$5.22 &279.15$\pm$6.89\\
$\left[NII\right] \lambda$6584    &  1.48$\pm$0.11 &  1.28$\pm$0.13 &  2.39$\pm$0.13 &2.10$\pm$0.16   &  0.87$\pm$0.02 &  0.81$\pm$0.03\\
HeI $\lambda$6678                 & 3.32$\pm$0.34  &  2.85$\pm$0.41 &  3.15$\pm$0.25 &2.75$\pm$0.31   &  3.47$\pm$0.15 &  3.23$\pm$0.20\\
$\left[SII\right] \lambda$6717    & 3.88$\pm$0.21  &  3.33$\pm$0.25 &  6.29$\pm$0.35 &5.47$\pm$0.43   &  1.64$\pm$0.07 &  1.52$\pm$0.09\\
$\left[SII\right] \lambda$6731    & 1.67$\pm$0.28  &  1.43$\pm$0.34 &  2.92$\pm$0.25 &2.54$\pm$0.31   &  0.70$\pm$0.07 &  0.65$\pm$0.09\\
                                  &       \\
F(H$\beta$)\footnote{In units of $\times$10$^{-14}$ erg cm$^{-2}$ s$^{-1}$} &0.66$\pm$0.01 & & 1.09$\pm$0.02 & & 1.91$\pm$0.03 \\
EW(H$\alpha$)\footnote{In units of $\rm \AA$}                               &970           & & 829           & &1311 \\ 
EW(H$\beta$)$^b$                                                            &272           & & 233           & &300 \\ 
c(H$\beta$)                                                                 & 0.21         & & 0.19          & &  0.10\\
\hline
\end{tabular}
\label{tb_lines}
\end{minipage}
\end{table*}
 
\subsection{Velocity field}\label{velocity_field}

\begin{figure}
\centering
\includegraphics[width=92mm]{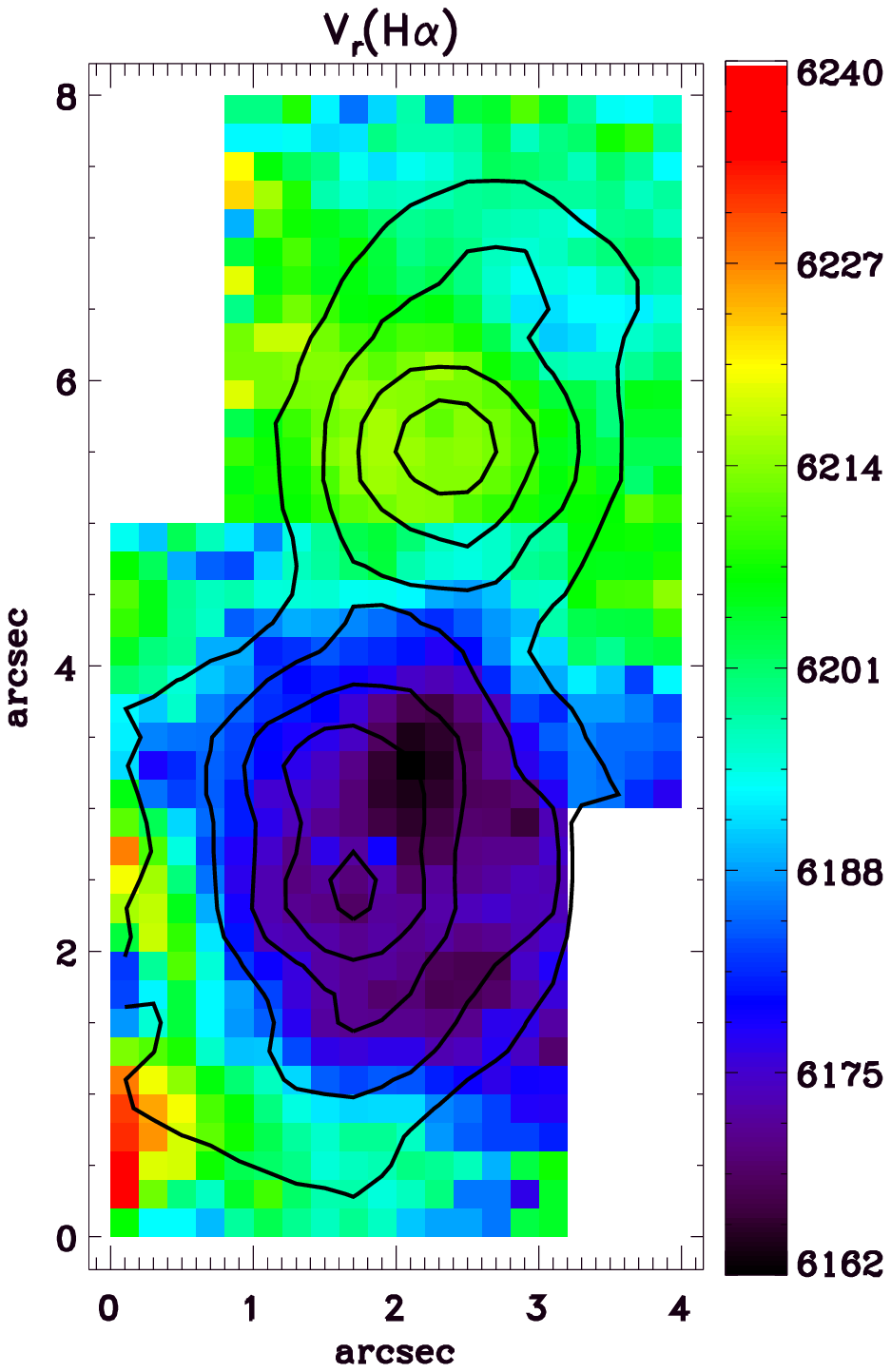}
 \caption{Recessional velocity of the H$\alpha$ emission line in units of km s$^{-1}$.
Contours display the H$\alpha$ morphology of HS\ 2236+1344.}
\label{velocidad}
\end{figure}

We have obtained the spatial distribution of radial velocities v$_{r}$ in the
ISM by fitting a single Gaussian to the H$\alpha$ emission line profiles. 
The velocity field of HS\ 2236+1344 is displayed in Fig.~\ref{velocidad}.
The v$_{r}$ displays a variation of $\sim$80~km s$^{-1}$ over the FoV, while in the
central part of the BCD it varies by about 45~km s$^{-1}$.
Additionally, the velocity field, in our FoV, close to the galaxy's center shows some rotation,
with its upper part (northwest; region 1) being redshifted and 
its lower part (southeast; regions 2 and 3) blueshifted with respect to the systemic galaxy velocity.
The relatively small measured variation in v$_{r}$ is in the range of values determined for other 
BCDs (e.g., van Zee et al. \cite{v01}, Cair\'os et al. \cite{Cairos12}) yet significantly larger 
than for the intrinsically faint BCD UM\ 408 (Lagos et al. \cite{L09}).

% ===================== Figure 4 ====================================================================

Moiseev et al. (\cite{M10}) studied the H$\alpha$ velocity field of HS\ 2236+1344
using the multi-mode Fabry-Perot instrument SCORPIO.
They concluded that, most likely, this galaxy is a strongly interacting 
pair of dwarf galaxies of comparable mass, possibly undergoing final stages of
merging. Additionally, they argue that the observed kinematics can not
be explained by giant ionized shells in a single rotating
disc, and, to the contrary, there is evidence for two independently rotating discs.
Our H$\alpha$ velocity field is similar to the one these authors determine, yet 
our data show no clear evidence for two kinematically distinct gas disks, or kinematical
perturbations indicative of a strongly interacting system.

% ===================== Figure 5 ====================================================================
\begin{figure}
\centering
\includegraphics[width=70mm]{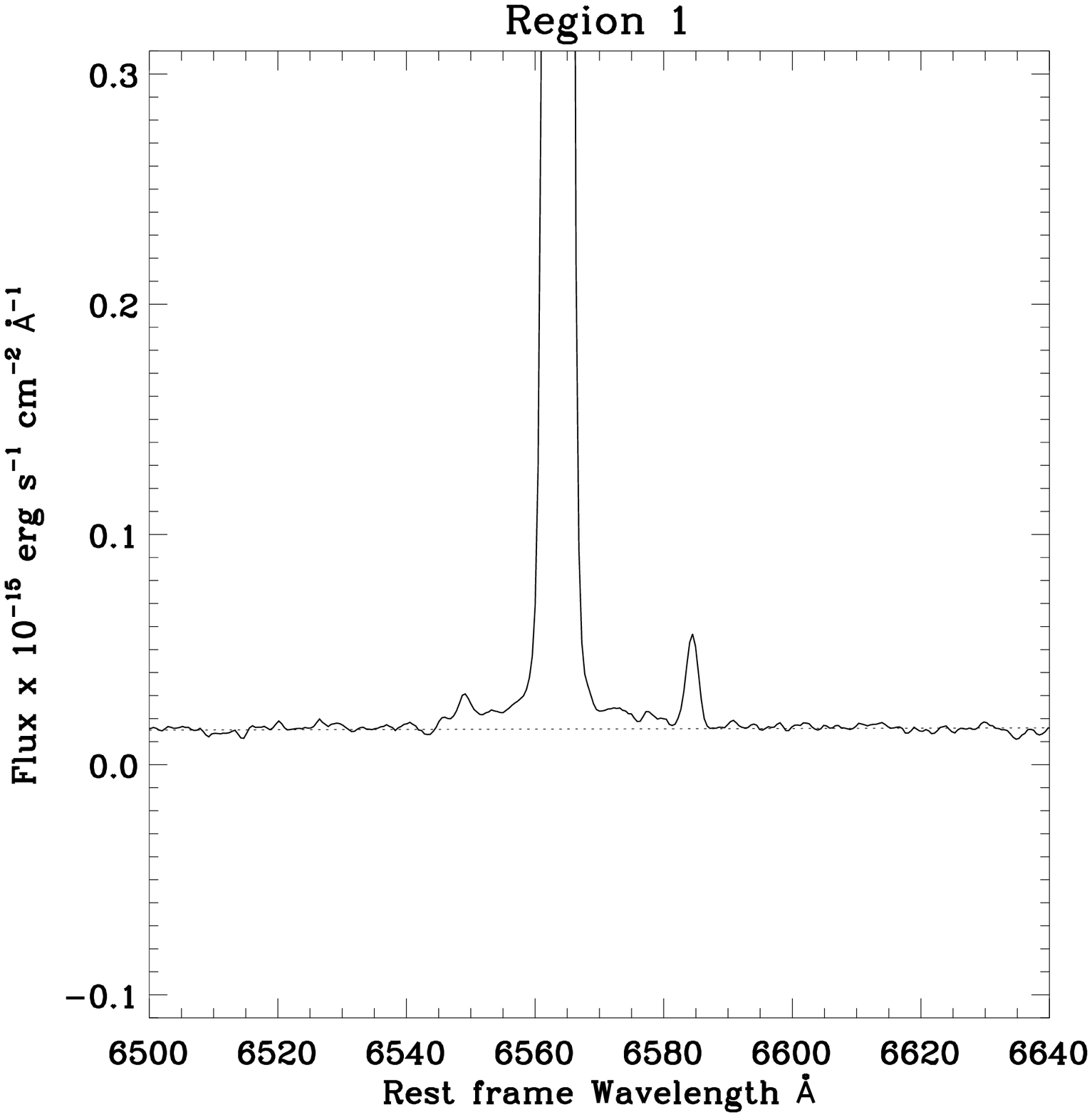}
\includegraphics[width=70mm]{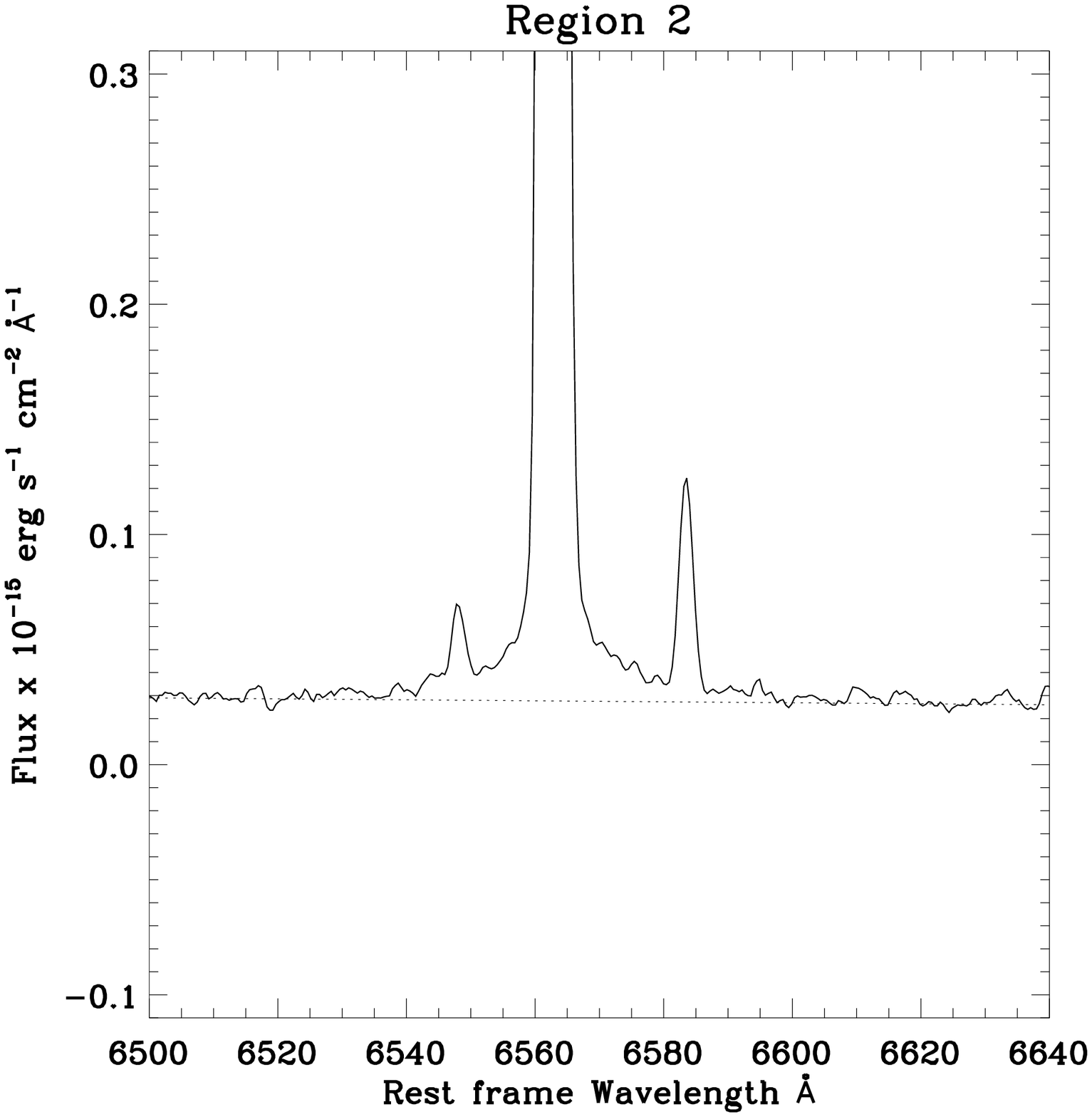}
\includegraphics[width=70mm]{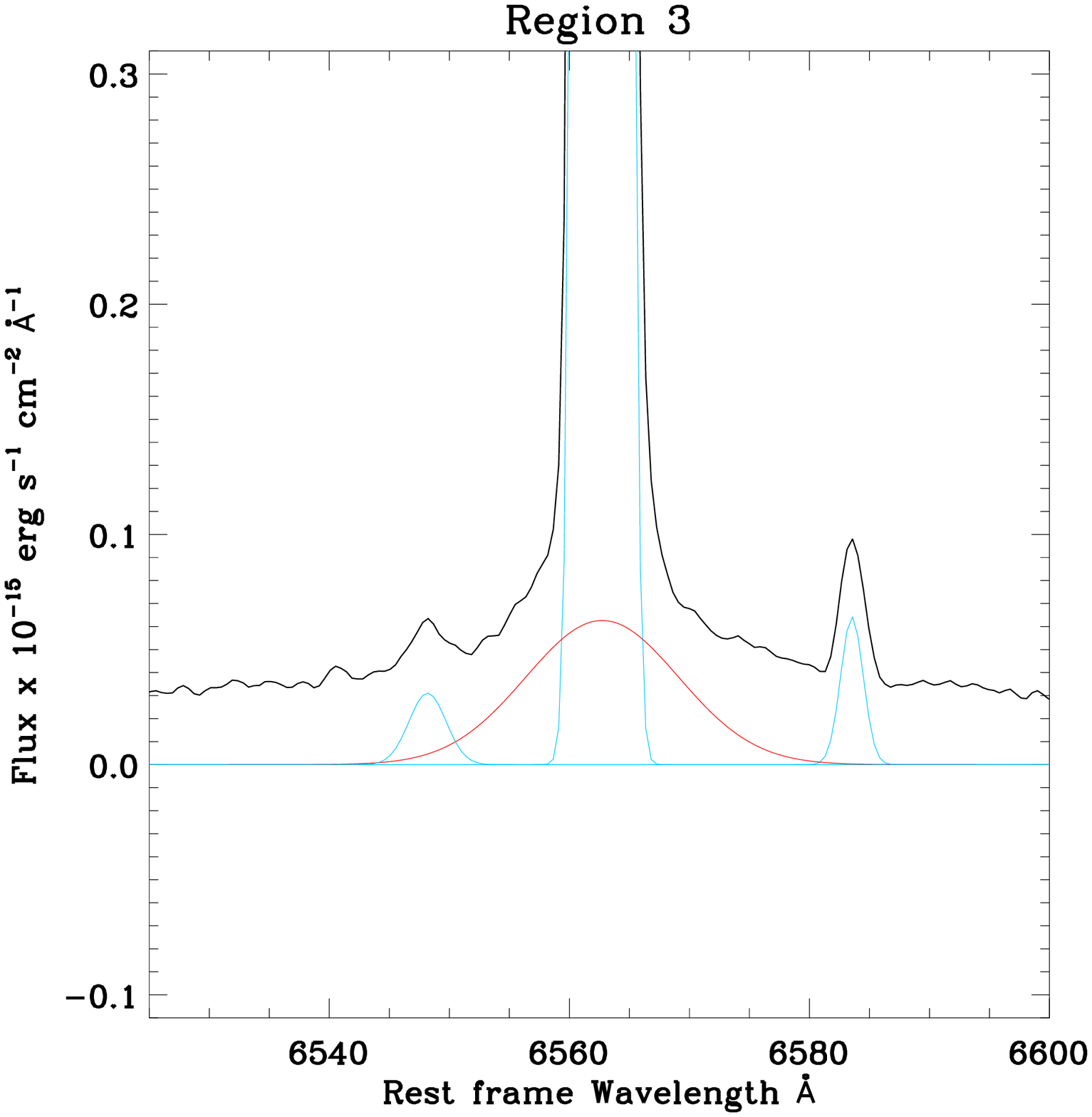}
\caption{H$\alpha$ emission line profiles of regions 1, 2 and 3, respectively.
Note the presence of broad emission (red line) in region 3.
}
\label{Ha_profile}
\end{figure}
 
We checked for the presence of a broad low-intensity component in the
H$\alpha$ emission line profiles (Fig.~\ref{Ha_profile}) using the integrated spectrum of the galaxy, as well as that
extracted from regions 1--3. For this, we have followed a similar procedure as that described in 
Lagos et al. (\cite{L12}):
We used the pan\footnote{http://www.ncnr.nist.gov/staff/dimeo/panweb/pan.html} 
routine (Peak ANalysis) in idl to fit two components to these profiles. 
Note that in different regions of the galaxy the H$\alpha$ emission line profile is relatively 
symmetric and well represented by a single Gaussian.
Although we use, in our study, the medium resolution grating R600, a detailed inspection of the base of the H$\alpha$ 
emission line profile (and other emission lines, such as [O \,{\sc iii}] $\lambda$5007) in regions 2 and 3 reveals 
a broad component at low intensity levels similar to the ones observed in other BCD galaxies 
(e.g., Izotov et al. \cite{Izotov1996}, Pustilnik et al. \cite{Pustilnik04}, Izotov \& Thuan \cite{IT09}).
We fit the narrow component of the H$\alpha$, [N\,{\sc ii}] $\lambda$6548 and [N\,{\sc ii}] $\lambda$6584  
by a single Gaussians (see Fig. \ref{Ha_profile}).
Then, simultaneously, we fit the broad component by a single Gaussian obtaining a FWHM of $\gtrsim$100$\rm \AA$.
A detailed analysis of this component is beyond the scope of this paper 
but its presence, in the inner part of region 3, appears to be consistent with 
stellar winds due by massive stars, e.g., from Luminous Blue Variables (LBV).

% ---------------------------------------------------------------
\subsection{The He\,{\sc ii} $\lambda$4686 emission line}\label{HeII}
% ---------------------------------------------------------------

\begin{figure}
\centering
\includegraphics[width=80mm]{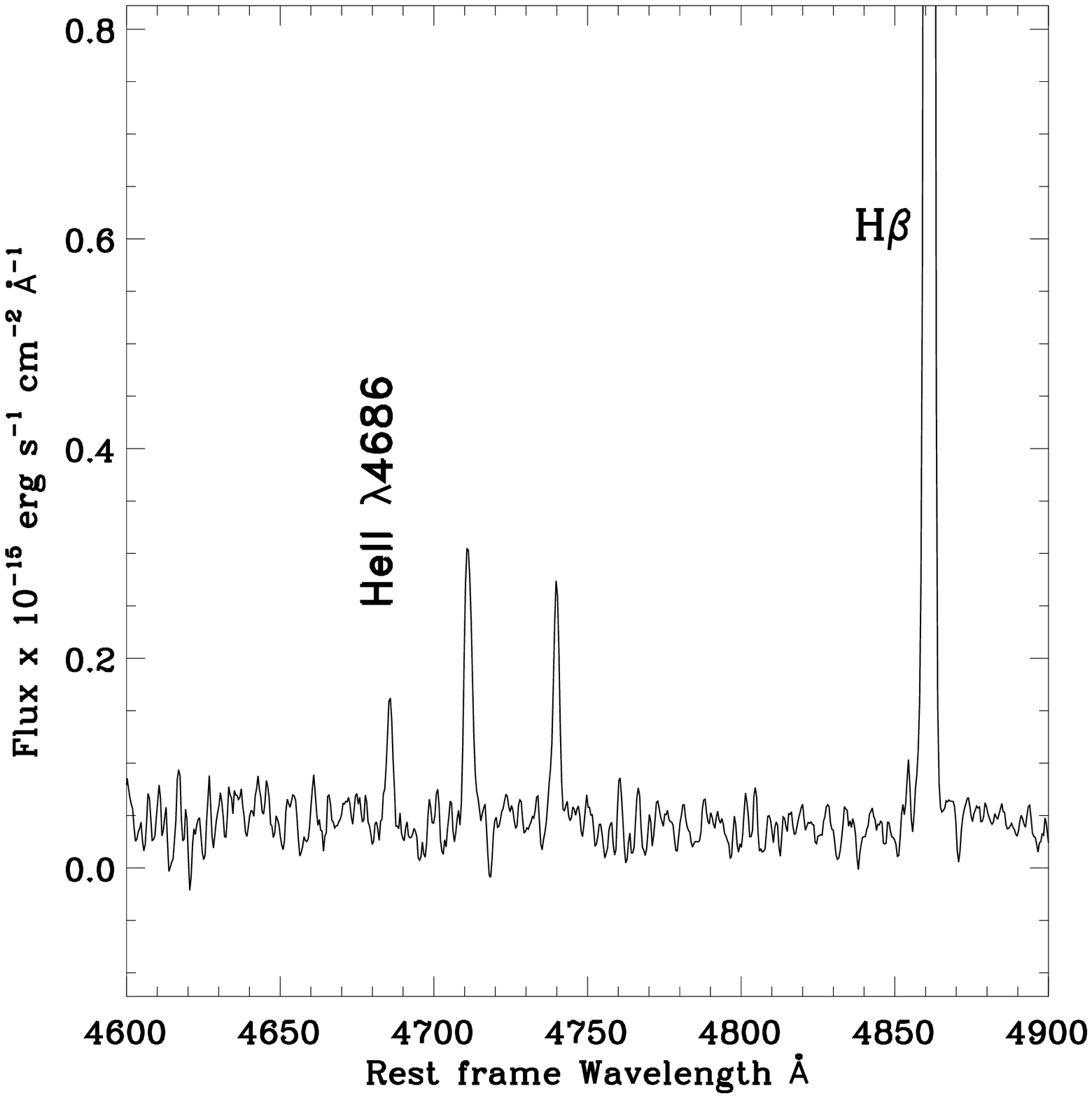}
\caption{Integrated spectrum of HS\ 2236+1344 between 4600 $\rm \AA$ to 4900 $\rm \AA$ 
with the He\,{\sc ii} $\lambda$4686 and H$\beta$ emission lines labeled. 
}
\label{R3_HeII}
\end{figure}

\begin{figure}
\centering
\includegraphics[width=90mm]{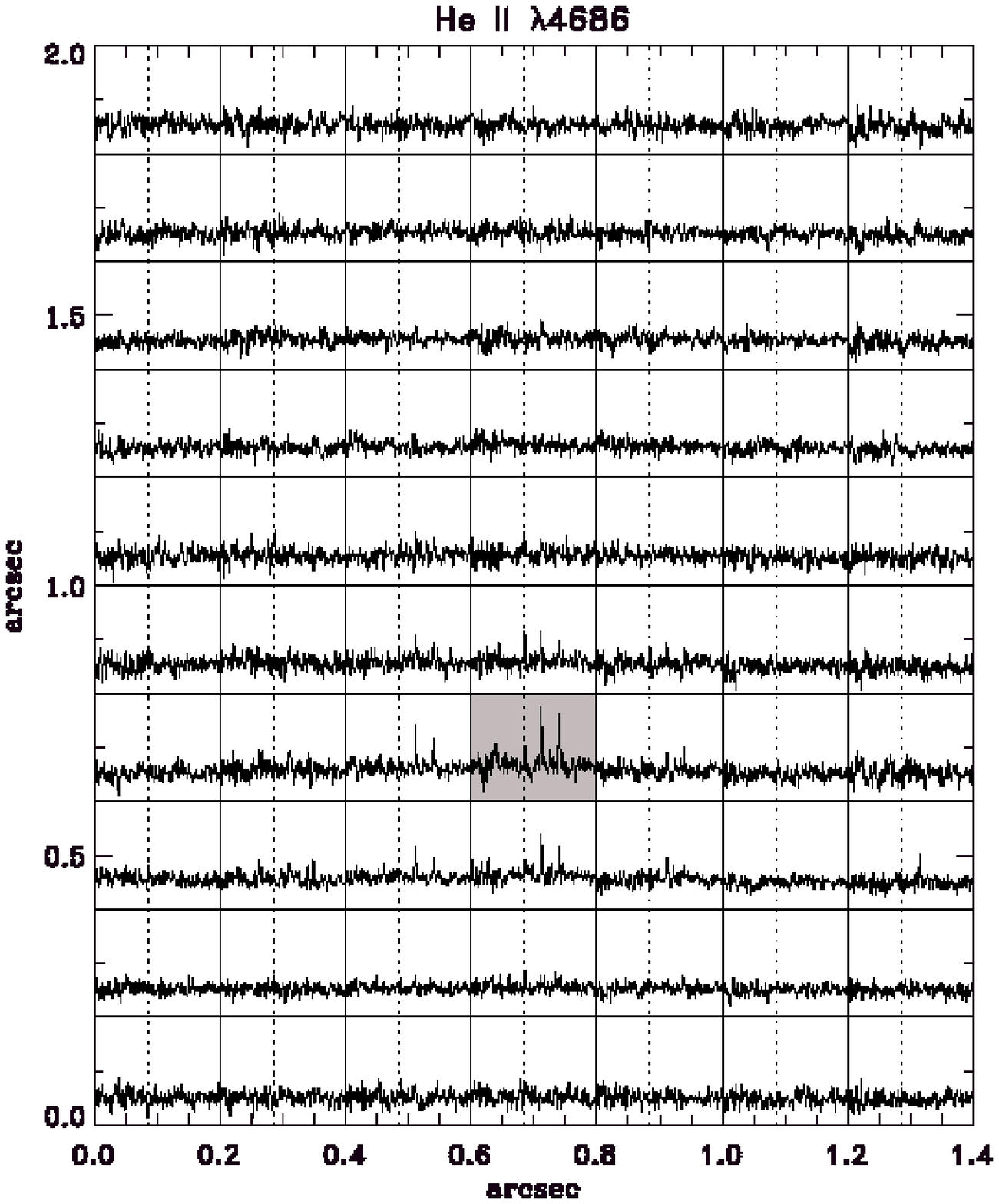}
 \caption{ 
Spectral region between 4600 $\rm \AA$ to 4800 $\rm \AA$ that includes the He\,{\sc ii} $\lambda$4686 
emission line in a rectangular area of the galaxy that contains regions 2 and 3.
Each square represents the area covered by an individual spaxel in that region.
The He\,{\sc ii} emission line in each spectrum
is indicated by the dotted line. 
The dark square corresponds to the H$\alpha$ peak.} 
\label{HeII_dist}
\end{figure}

In Fig.~\ref{R3_HeII} we show the integrated spectrum of region 3 in the wavelength range 
from 4600 $\rm \AA$ to 4900 $\rm \AA$, which reveals the presence of a narrow He\,{\sc ii} $\lambda$4686
emission.
Figure~\ref{HeII_dist}, encompassing regions 2 and 3, allows a study of the spatial distribution
of that line. Squares represent the area on the sky that is covered by individual spaxels
and are overlaid with the respective observed spectrum in the wavelength range 
4600 $\rm \AA$ -- 4800 $\rm \AA$.
The rest-frame wavelength of the emission line He\,{\sc ii} $\lambda$4686 is marked by dotted vertical lines,
and the dark square indicates the location of the peak H$\alpha$ intensity.

% ===================== Figure 6 ====================================================================

It is apparent from Fig.~\ref{HeII_dist} that the high-ionization He\,{\sc ii} $\lambda$4686 emission is not extended
as, e.g., in the compact H\,{\sc ii}/BCD galaxies \object{Tol~2146-391} (Lagos et al. \cite{L12}) and Mrk~178 
(Kehrig et al. \cite{Kehrig13}). On the contrary, we find that most of the He\,{\sc ii} $\lambda$4686 emission 
in HS\ 2236+1344 is confined within region~3 (see Fig.~\ref{HeII_dist}),
near the peak of H$\alpha$ emission. This is presumably due to the low S/N in the blue part 
of our GMOS spectra, which likely prevents detection of extended low-surface brightness 
in He \,{\sc ii} emission that may be present. This also suggested by the fact that previous high-S/N long slit 
spectroscopy (Izotov \& Thuan \cite{IT07}) has revealed He\,{\sc ii} emission in region \#1 as well. 
In any case, the spatial association of He\,{\sc ii} $\lambda$4686 with region \#3 suggests that 
its excitation is due to the ongoing burst.
We note that the intensity I(He\,{\sc ii} $\lambda$4686)/I(H$\beta$) =
0.0122$\pm$0.0029 in region 3 agrees within the errors with
I(He\,{\sc ii} $\lambda$4686)/I(H$\beta$)=0.0105$\pm$0.0008 and 0.0112$\pm$0.0013 obtained by
Izotov \& Thuan (\cite{IT07}) for their regions 1 and 2, respectively.

% ===================== Figure 7 ====================================================================

One of the most likely explanations\footnote{See, e.g., Thuan \& Izotov (\cite{TI05}) and references therein 
for other candidate sources for He\,{\sc ii} $\lambda$4686 ionization.} 
for the high ionization emission lines in BCD
galaxies is the presence of fast radiative shocks (Dopita \& Sutherland \cite{DS96}) in the ISM of these galaxies 
(e.g., Thuan \& Izotov \cite{TI05}, Izotov et al. \cite{I06}, Lagos et al. \cite{L12}).  
As noted above, we do not detect the He\,{\sc ii} $\lambda$4686 emission line 
at the arm-like structures in HS\ 2236+1344 or within any other extended morphological feature. 
On the other hand, at ages of $\sim$3-6 Myr the WR phase is still ongoing (Leitherer et al. \cite{L99}), 
so we expect a significant amount of these stars in the young regions (with ages of $\sim$3 Myr; 
see Sect. \ref{sf_io_mech} for more details) found in this galaxy.
But, we did not detect clear WR features at $\lambda$4686 $\rm \AA$ 
neither in the integrated spectrum of the galaxy nor in the different regions resolved in HS\ 2236+1344.
Although, in the case of WR stars the He\,{\sc ii} $\lambda $4686 emission line 
and the blue WR bump are mainly linked to WN stars, the non-detection of this stellar emission
suggests the existence of WR stars in an early evolutionary stage (e.g., LBV stars).
In fact, a large fraction of galaxies with oxygen abundances lower than 12+log(O/H)=8.2
do not show WR features (Guseva et al. \cite{Guseva00}, Brinchmann et al. \cite{Brinchmann08}, 
Shirazi \& Brinchmann \cite{SB12}).
It is also known, that in several H\,{\sc ii}/BCDs (e.g., \object{II Zw 70} and Mrk 178; 
Kehrig et al. \cite{K08,Kehrig13}) He\,{\sc ii} $\lambda $4686 emission does not strictly coincide 
with the location of the WR bumps.
In this sense, He\,{\sc ii} emission without associated WR features does not completely 
rule out a contribution of WR stars to the He\,{\sc ii} ionization, even though other excitation sources 
are likely dominating. 

In summary, various ionization sources may appear consistent with
the observed morphology of the He\,{\sc ii} $\lambda$4686 emission in HS\ 2236+1344, whereas  
from the present data this seems to be an unresolved issue.

\subsection{Emission Line Ratios}\label{diagramas}

Using the information derived in the previous sections
it is possible to attempt distinguishing among different ionization
mechanisms using BPT (Baldwin, Phillips and Terlevich; Baldwin et al. \cite{B81}) diagrams.
To do this, we adopt the following emission line ratios:
[O\,{\sc iii}] $\lambda$5007/H$\beta$, [N\,{\sc ii}] $\lambda$6584/H$\alpha$, 
[S \,{\sc ii}] $\lambda\lambda$6717, 6731/H$\alpha$ and [O\,{\sc i}] $\lambda$6300/H$\alpha$.
The [O\,{\sc iii}] $\lambda$5007/H$\beta$ emission
line ratio is an excitation indicator and provides
information about the available fraction of hard ionizing photons in an H\,{\sc ii} region.
On the other hand, [N\,{\sc ii}] $\lambda$6584, [S \,{\sc ii}] $\lambda\lambda$6717, 6731 and 
[O\,{\sc i}] $\lambda$6300 are low-ionization emission lines, usually weak in H\,{\sc ii} regions. 
So, the emission line ratios
[N\,{\sc ii}] λ6584/H$\alpha$, [S \,{\sc ii}] $\lambda\lambda$6717, 6731/H$\alpha$ 
and [O\,{\sc i}] $\lambda$6300/H$\alpha$ can effectively disentangle between
photoionization under physical conditions that are typical for H\,{\sc ii}
regions and other excitation mechanisms (e.g., AGN or shocks).

\begin{figure}
\centering
\includegraphics[width=42mm]{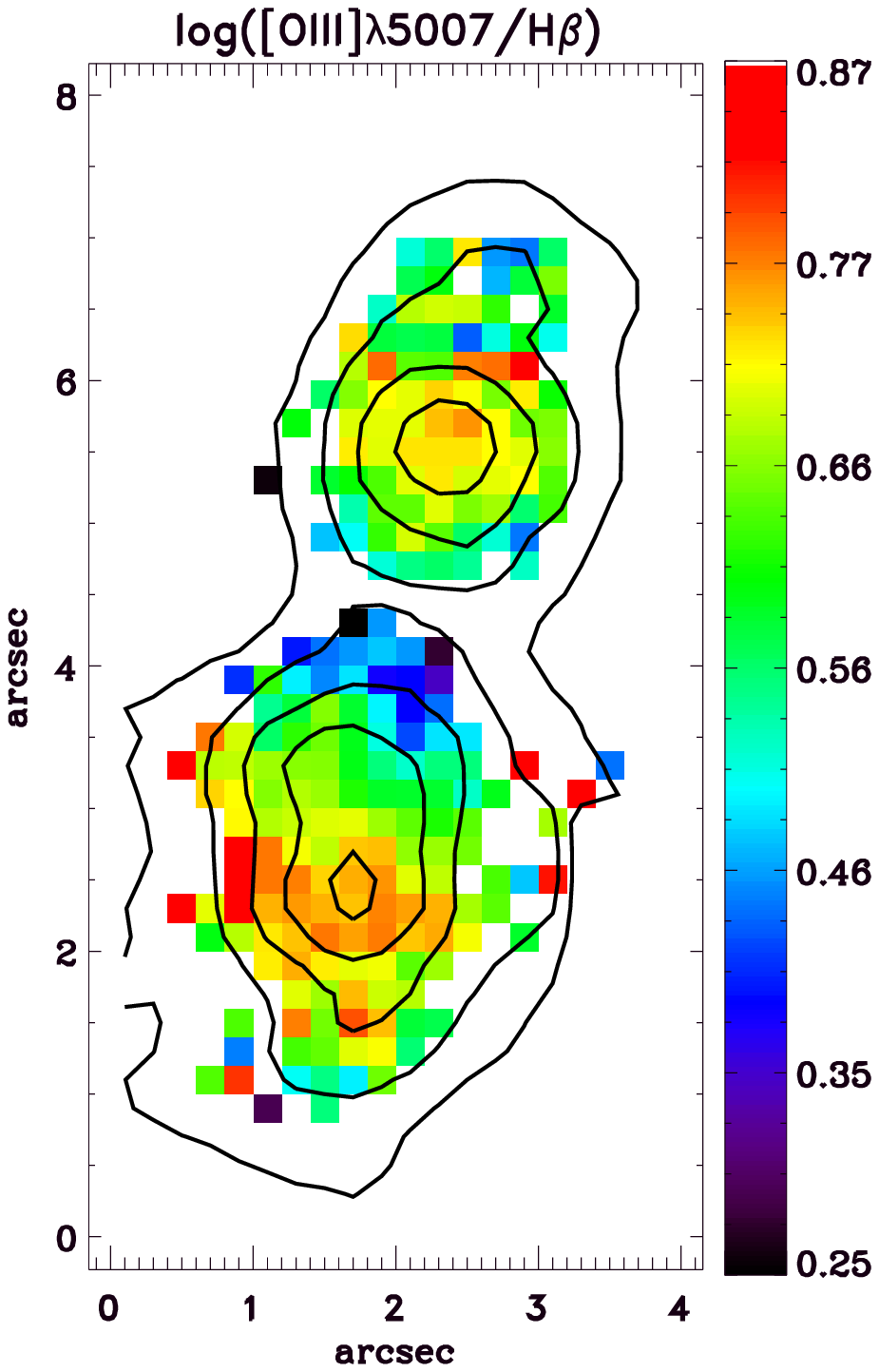}
\includegraphics[width=42mm]{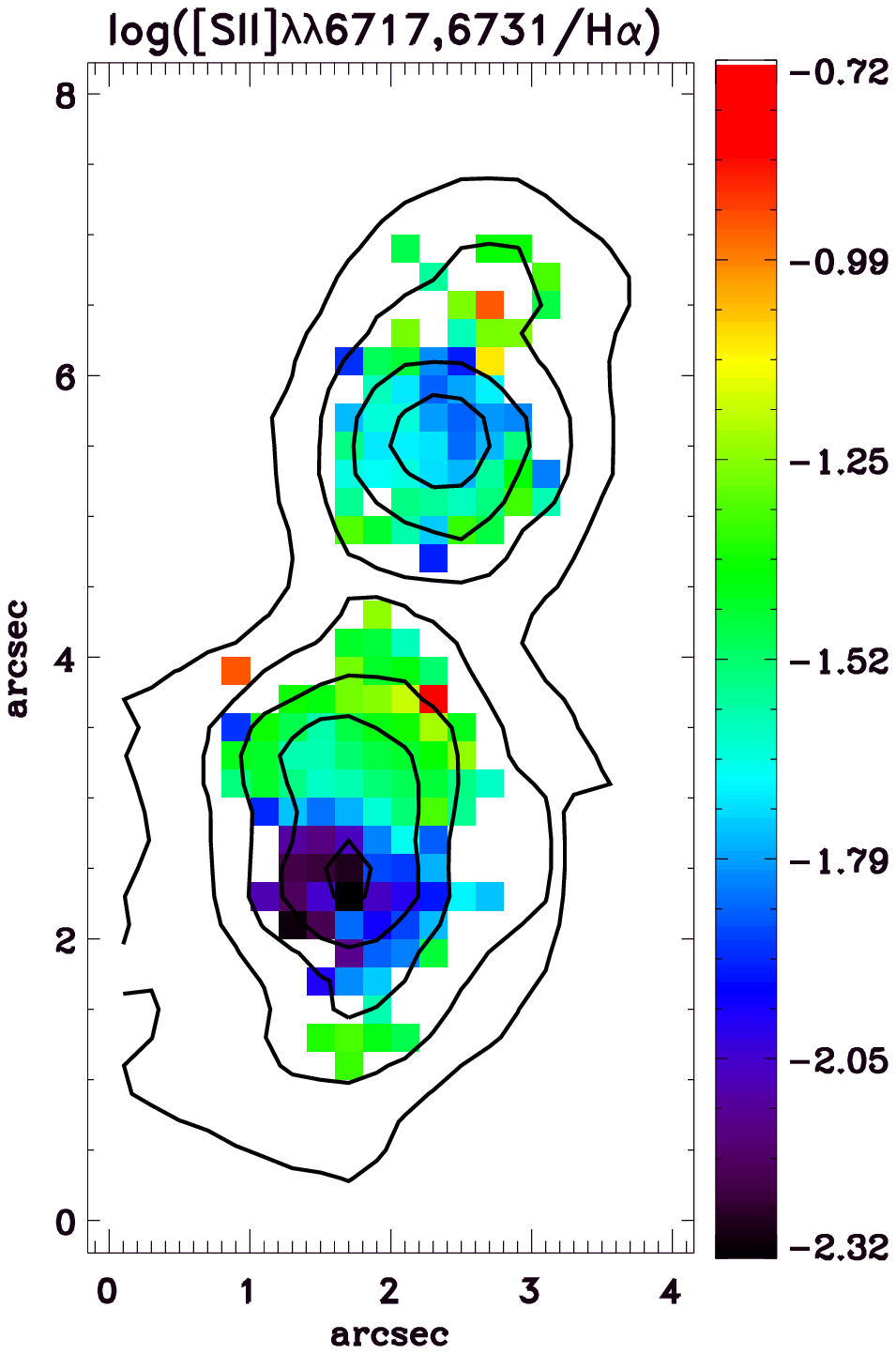}\\
\includegraphics[width=42mm]{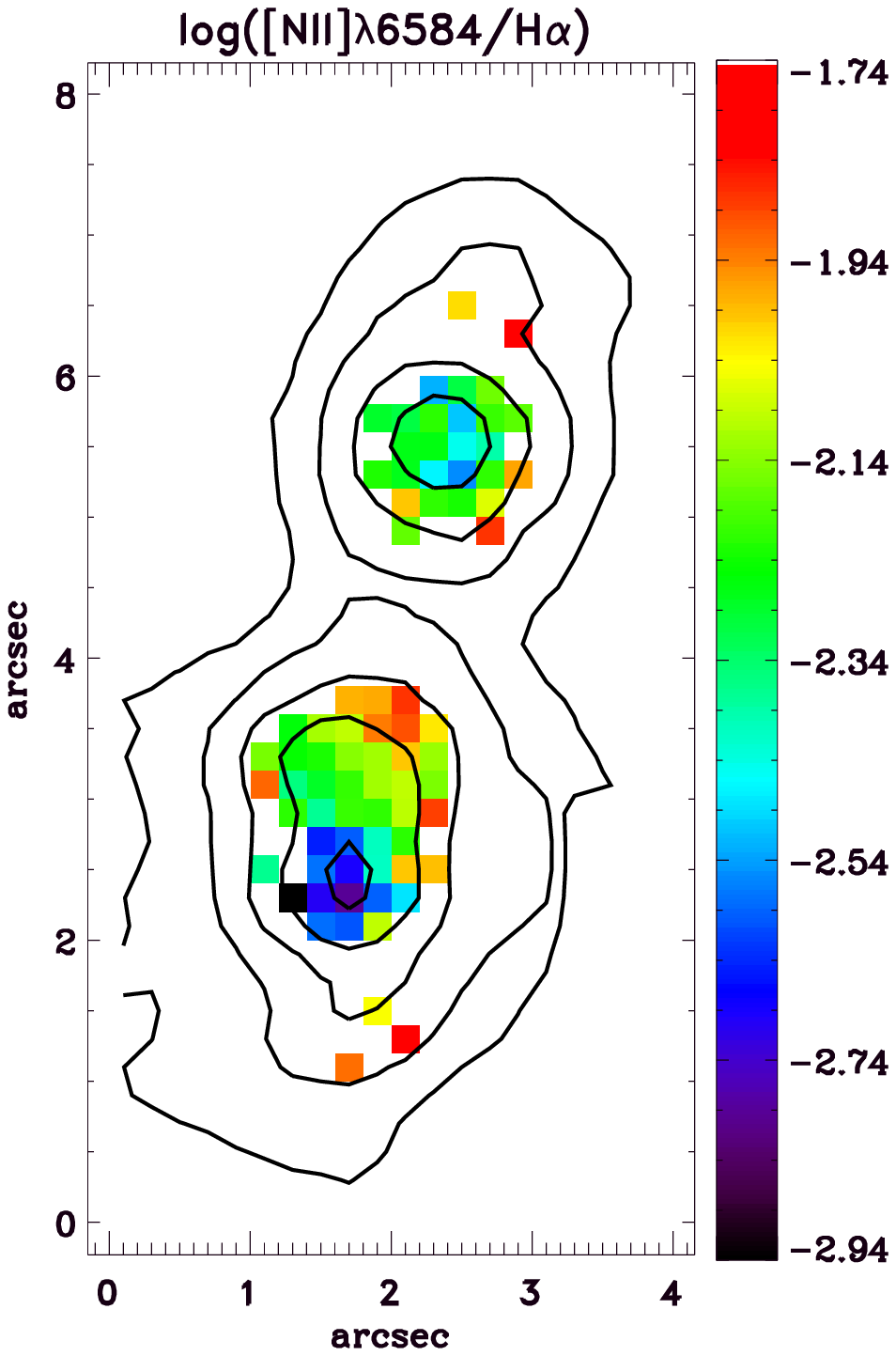}
\includegraphics[width=42mm]{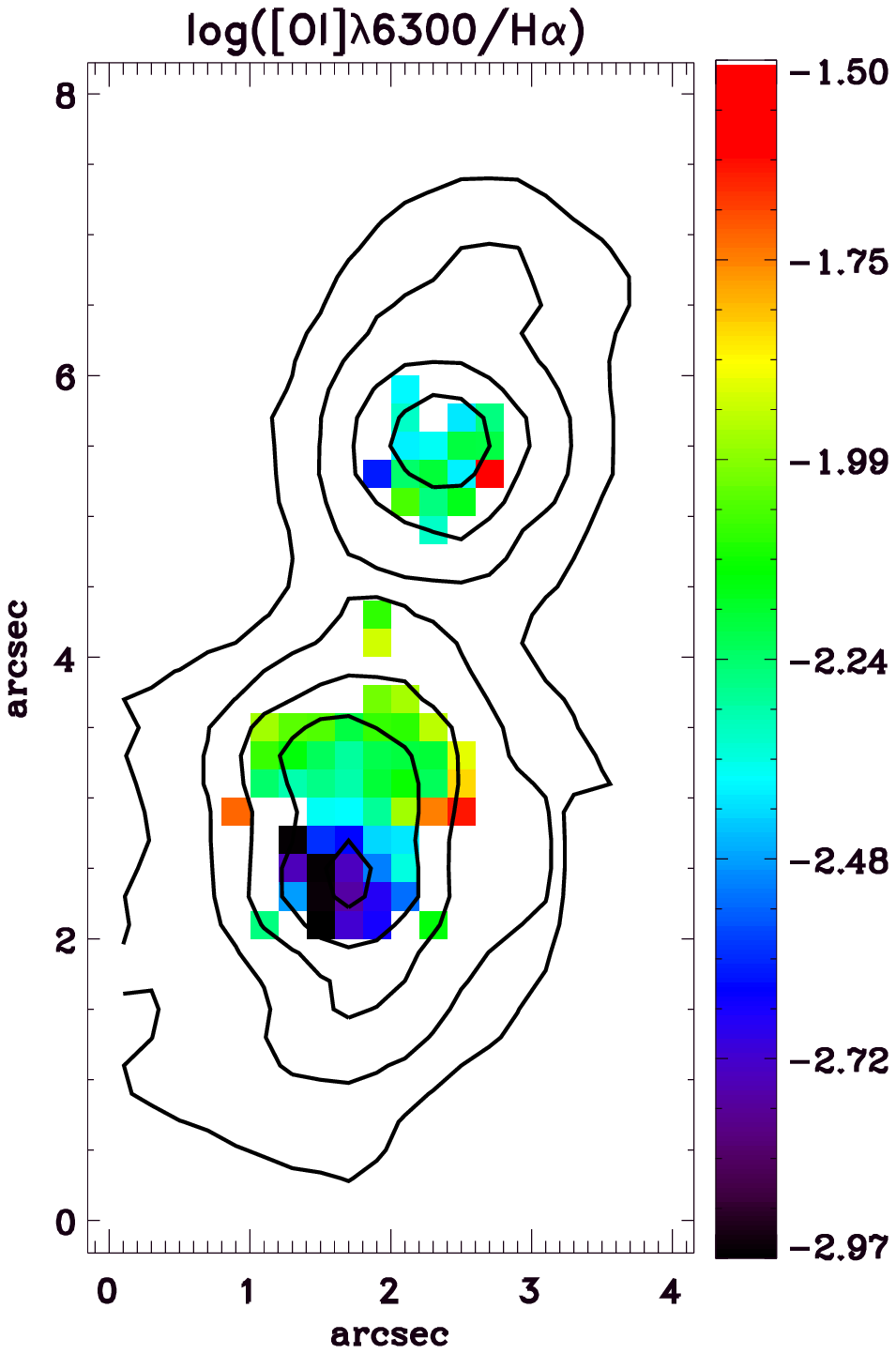}
 \caption{Emission line ratios: log [O\,{\sc iii}]$\lambda$5007/H$\beta$, 
log [S\,{\sc ii}]$\lambda\lambda$6717,6731/H$\alpha$, log [N\,{\sc ii}]$\lambda$6584/H$\alpha$ 
and [O\,{\sc i}]$\lambda$6300/H$\alpha$. Overlaid are the H$\alpha$ flux contours. 
}
\label{emi_ratios_spa}
\end{figure}

In Fig.~\ref{emi_ratios_spa} we show the emission line ratio maps   
[O\,{\sc iii}] $\lambda$5007/H$\beta$, [N\,{\sc ii}] $\lambda$6584/H$\alpha$, 
[S\,{\sc ii}] $\lambda\lambda$6717,6731/H$\alpha$ and [O\,{\sc i}] $\lambda$6300/H$\alpha$, respectively. 
In this figure, the ionization structure in the central part of regions 1, and 3 of 
HS\ 2236+1344 is rather uniform for all emission line ratios, 
while the emission line ratio [O\,{\sc iii}]~$\lambda$5007/H$\beta$ decreases slightly, in region 2, with 
distance. The opposite is observed for the [S\,{\sc ii}] $\lambda\lambda$6717,6731/H$\alpha$, 
[N\,{\sc ii}] $\lambda$6584/H$\alpha$ and [O\,{\sc i}] $\lambda$6300/H$\alpha$ ratios.

% ===================== Figure 8 ====================================================================

\begin{figure}
\centering
\includegraphics[width=90mm]{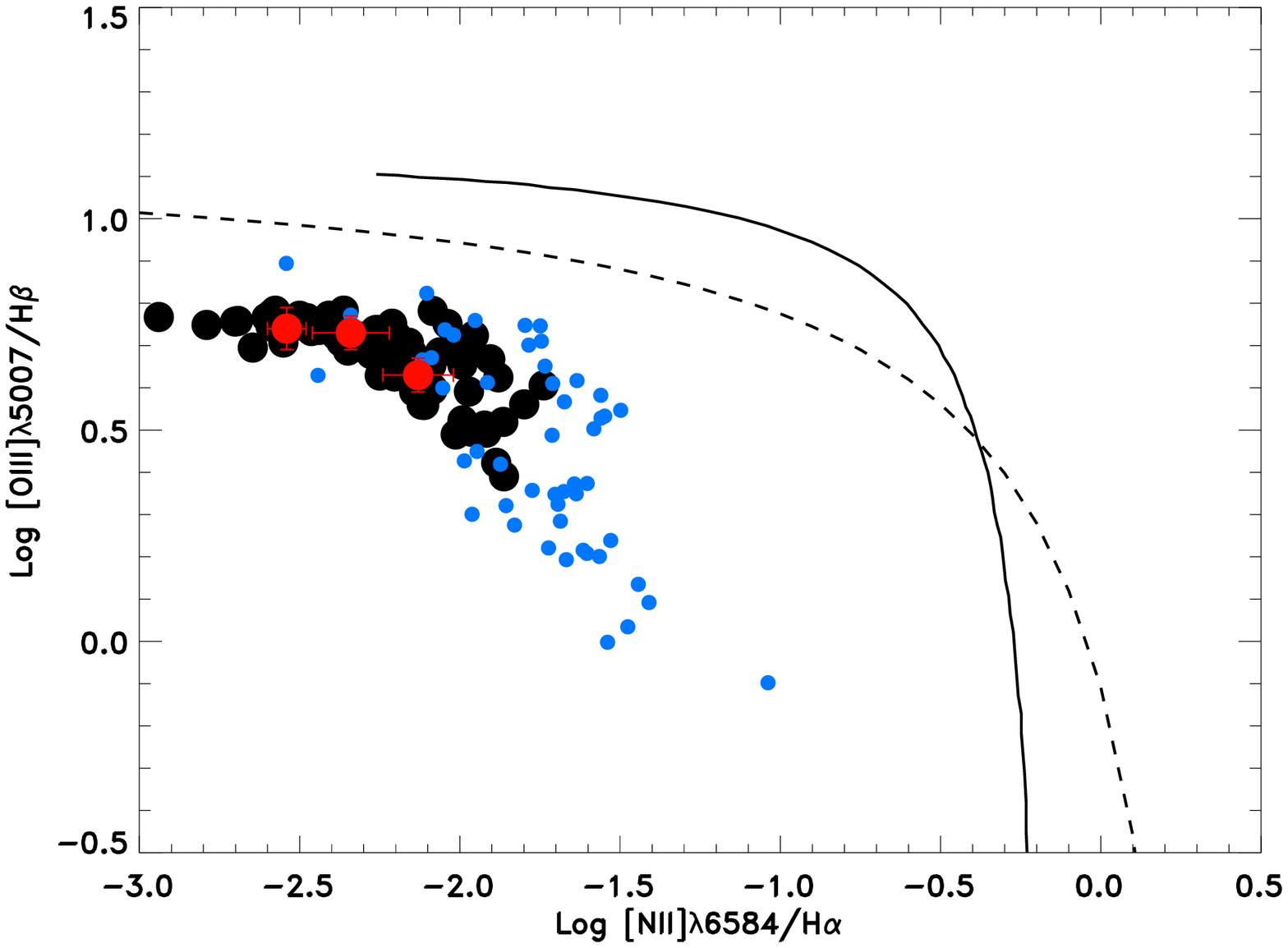}
\caption{log [O\,{\sc iii}]$\lambda$5007/H$\beta$ 
versus log [N\,{\sc ii}]$\lambda$6584/H$\alpha$ BPT diagram. 
The solid curves show the empirical borders between photoionization by massive stars and other excitation mechanisms
(Osterbrock \& Ferland \cite{O06}), while the dotted lines show the demarcation proposed by Kewley et al. (\cite{K01}).
Black dots correspond to individual spaxels in Fig.~\ref{emi_ratios_spa}.
The values for region 1, 2 and 3 of HS\ 2236+1344 are indicated in the diagram by red points.
Blue data points show values obtained by Izotov et al. (\cite{I12}).
}
\label{BPT_galaxias}
\end{figure}

In Figure~\ref{BPT_galaxias} we show the BPT diagram comparing [O\,{\sc iii}]$\lambda$5007/H$\beta$
versus [N\,{\sc ii}] $\lambda$6584/H$\alpha$. In this figure, the solid line 
(adapted from Osterbrock \& Ferland \cite{O06}) shows the demarcation between H\,{\sc ii} regions photoionized 
by massive stars and regions dominated by other excitation sources.
The dotted line represents the same as the solid line but using the models given by Kewley et al. (\cite{K01}). 
Black dots correspond to the values of individual spaxels 
in Fig.~\ref{emi_ratios_spa}. Red dots show the integrated values of regions 1, 2 and 3, while the blue ones
are several low metallicity (12+log(O/H)$\lesssim$7.6) SF regions from Izotov et al. (\cite{I12}).
It can be seen from Fig.~\ref{BPT_galaxias} that the data points of HS\ 2236+1344 occupy the
left region of the BPT diagram, showing smaller [N\,{\sc ii}] $\lambda$6584/H$\alpha$ ratios, 
while the [O\,{\sc iii}] $\lambda$5007/H$\beta$ ratio increases.  
Therefore, the data points of HS\ 2236+1344 occupy the same region of the diagram, 
the left part of the BPT diagram, than other XMP galaxies found in the literature (Izotov et al. \cite{I12}). 

% ===================== Figure 9 ====================================================================

Finally, we note that our integrated [O\,{\sc iii}] $\lambda$5007/H$\beta$, 
[S\,{\sc ii}] $\lambda\lambda$6717,6731/H$\alpha$, 
[N\,{\sc ii}] $\lambda$6584/H$\alpha$ and [O\,{\sc i}] $\lambda$6300/H$\alpha$ values 
for HS\ 2236+1344 (Table \ref{integrated_properties_1}) are consistent with the ones derived by  
Thuan \& Izotov (\cite{TI05}) of 0.68, -1.65, -2.21 and -2.39, respectively.

\subsection{Chemical abundances} \label{chemical}

In order to determine the chemical abundance pattern in the ISM of HS\ 2236+1344, 
we used the five level atomic model FIVEL (De Robertis et al. \cite{D87}), implemented in the 
IRAF STSDAS package $\it{nebular}$. 
We first calculated the electron temperature T$_{e}$(O\,{\sc iii})
from the ratio [O\,{\sc iii}] $\lambda\lambda$4959,5007/[O\,{\sc iii}] $\lambda$4363 
and the electron density from the ratio [S\,{\sc ii}] $\lambda$6717/[S\,{\sc ii}] $\lambda$6731.
The latter is typically $>$1, which indicates a low electron density n$_{e}$ 
(Osterbrock \& Ferland \cite{O06}). 
Large [S\,{\sc ii}] $\lambda$6717/[S\,{\sc ii}] $\lambda$6731 ratios were already 
observed in other studies (e.g., Lagos et al. \cite{L09}, L\'opez-Hern\'andez et al. \cite{LH13}, 
Krabbe et al. \cite{Krabbe14}). These nonphysical values could be partly due to associated 
with uncertainties in the measurements of these emission lines, due to the placement of the continuum or, 
more likely, because the spectrum was not corrected for night sky absorption lines. 
Therefore, intensities of [S\,{\sc ii}] $\lambda$6717 and [S\,{\sc ii}] $\lambda$6731 
(Tables \ref{tb_lines_integrated} and \ref{tb_lines}) lines are, likely, affected by the night sky absorption.
In any case, we compare our results with the SDSS spectrum of the same galaxy obtaining 
[S\,{\sc ii}] $\lambda$6717/[S\,{\sc ii}] $\lambda$6731 $\sim$1.6, which is about 20\% larger than the value 
obtained by Thuan \& Izotov (\cite{TI05}) of 1.31 ($\sim$110 cm$^{-3}$). 
Therefore, it is reasonable to assume a value of 100 cm$^{-3}$ for all apertures in our calculations.

\begin{figure}
\centering
\includegraphics[width=90mm]{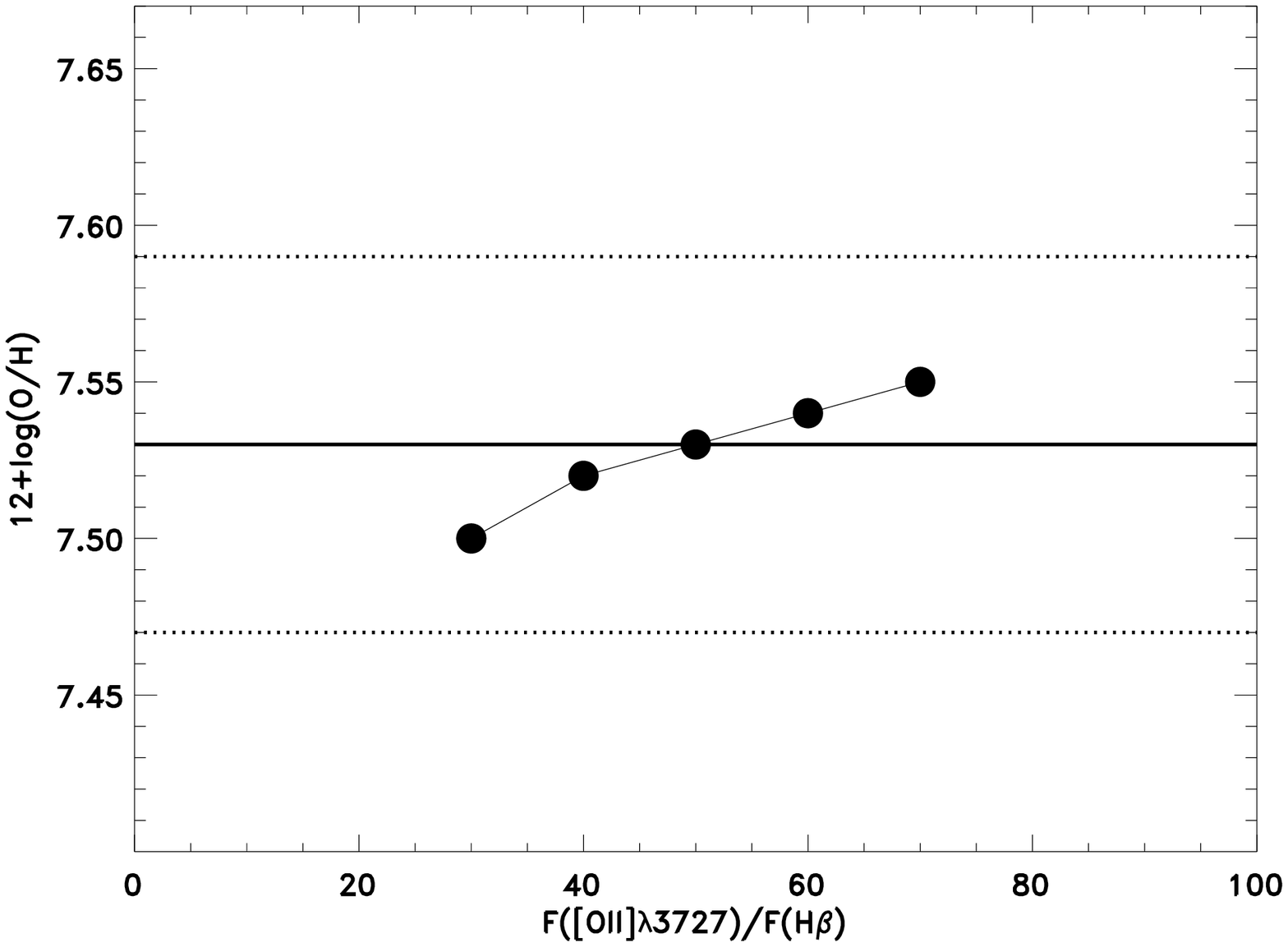}
 \caption{Variation of the 12+log(O/H) abundance in the integrated galaxy (black data points) 
 as a function of the adopted F([O{\sc ii}] $\lambda$3727)/F(H$\beta$) ratio. 
 The fluxes are relative to F(H$\beta$)=100.
 The continuum lines corresponds to the integrated values of 7.53 and 
 the dotted lines show the uncertainties associated with these values. More details are in the text.
}
\label{abundance_variation}
\end{figure}

The oxygen and nitrogen abundances were calculated as 
\begin{equation}
\frac{O}{H}=\frac{O^{+}}{H^{+}} + \frac{O^{+2}}{H^{+}},
\end{equation}
and
\begin{equation}
\frac{N}{H}=ICF(N) \frac{N^{+}}{H^{+}},
\end{equation}

\noindent
with the ionic abundances O$^{+}$, O$^{+2}$ and N$^{+}$ obtained from 
the \textit{nebular} output file. Nitrogen abundances were calculated using  
an ionization correction factor (ICF) ICF(N)=(O$^{+}$+O$^{+2}$)/O$^{+}$. 
We assumed that the T$_{e}$(O\,{\sc ii}) temperature is given by 
T$_{e}$(O\,{\sc ii}) = 2/(T$_{e}^{-1}$(O\,{\sc iii})+0.8) (Pagel et al. \cite{P92}) and 
T$_{e}$(O\,{\sc ii}) = T$_{e}$(N\,{\sc ii}). 
Since the [O\,{\sc ii}] $\lambda$3727 doublet is not within the spectral range of our data, 
we have to rely on published fluxes.
We therefore adopted ratios of F(O\,{\sc ii} $\lambda$3727)/F(H$\beta$)=54.67 (Thuan \& Izotov \cite{TI05}) 
for the integrated spectrum, 40.33$\pm$0.69 for region 1 and 25.39$\pm$0.41 for regions 2 and 3 
(Izotov \& Thuan \cite{IT07}), respectively. The fluxes are relative to F(H$\beta$)=100.
We checked our results with an alternative method for deriving abundances in which a constant
F(O\,{\sc ii} $\lambda$3727)/F(H$\beta$) value is assumed (Thuan \& Izotov \cite{TI05}). 
We found that values derived using a constant F(O\,{\sc ii} $\lambda$3727)/F(H$\beta$) ratio 
yield oxygen abundances $\sim$0.03 dex higher and nitrogen abundances $\sim$0.23 dex lower
than the ones obtained using the values of Izotov \& Thuan (\cite{IT07}). 
In Fig.~\ref{abundance_variation} we show the variation of the 12+log(O/H) abundance in the integrated galaxy 
as a function of the adopted F([O{\sc ii}] $\lambda$3727)/F(H$\beta$) ratio.
From this figure we can see that very small changes were found by adopting the
values by Thuan \& Izotov (\cite{TI05}) and Izotov \& Thuan (\cite{IT07}), since the oxygen abundance 
do not vary strongly with F(O\,{\sc ii} $\lambda$3727)/F(H$\beta$) within the errors.
So, the adoption of [O\,{\sc ii}] $\lambda$3727 fluxes from 
the literature introduces only minor uncertainties in the analysis of 12+log(O/H), while 
the N/O abundance ratio is highly uncertain.

% ===================== Figure 10 ====================================================================

\begin{figure*}
\centering
\includegraphics[width=65mm]{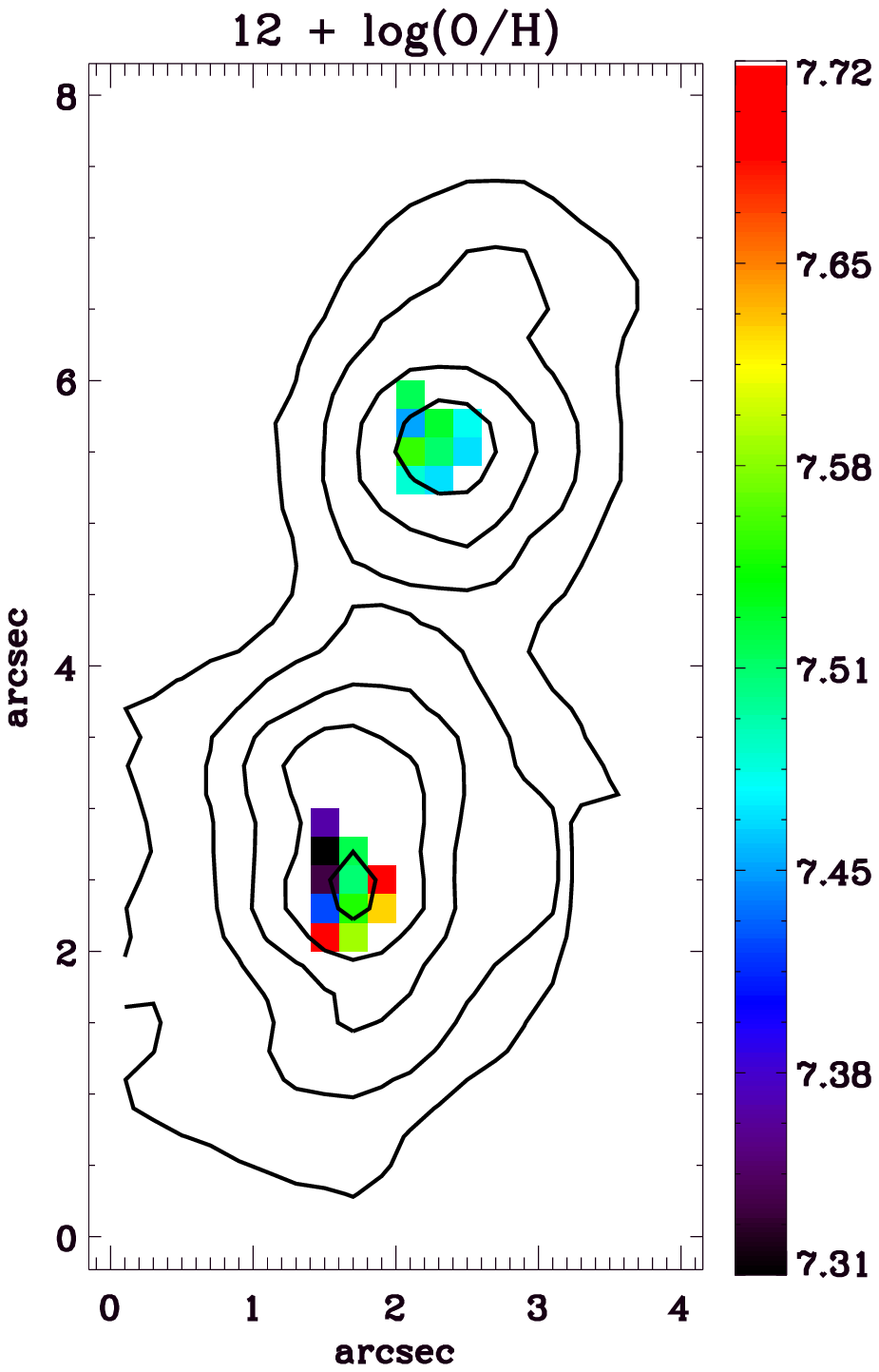}
\includegraphics[width=65mm]{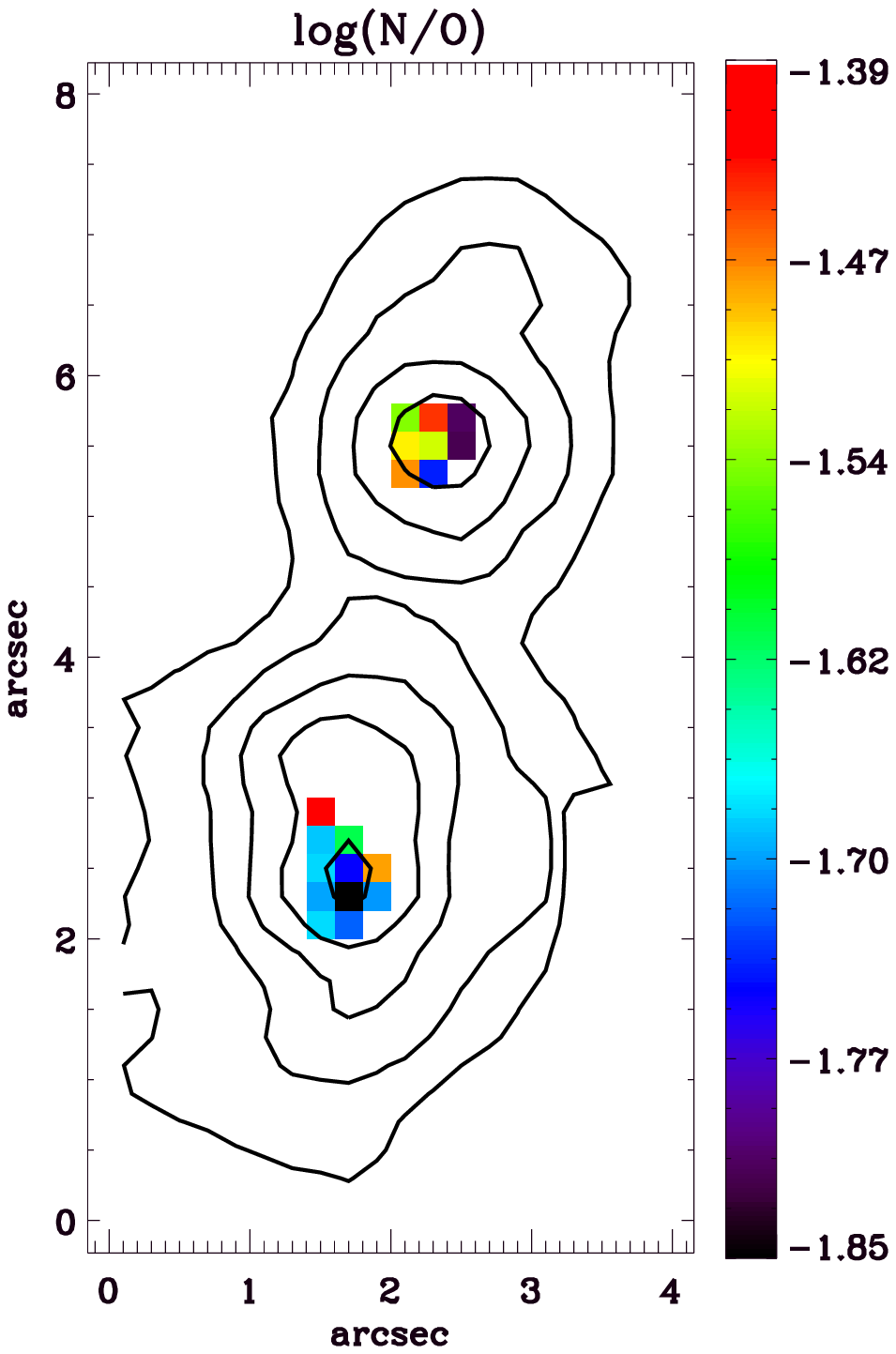}
 \caption{12+log(O/H) abundance and log(N/O) ratio maps in spaxels with determined T$_e$(O\,{\sc iii}). 
Overlaid are the H$\alpha$ flux contours.
}
\label{abundance_maps}
\end{figure*}

In Table \ref{integrated_properties_1}  
we show the electron density, electron temperature and abundances 
calculated for each one of the integrated apertures considered in this work. 
Figure~\ref{abundance_maps} (left panel) shows the spatial distribution 
of the oxygen abundances in spaxels with detected [O\,{\sc iii}] $\lambda$4363 emission,
and in the right panel of the same figure we show the
spatial distribution of log(N/O) calculated using the oxygen and nitrogen maps 
in the spaxels with detected [N\,{\sc ii}] $\lambda$6584 emission.

% ===================== Figure 11 ====================================================================

% ===================== Table 5 ====================================================================
\begin{table*}
 \centering
 \begin{minipage}{110mm}
  \caption{Ionic abundances and integrated properties of HS\ 2236+1344.}
  \begin{tabular}{@{}lcccc@{}}
\hline
                                                        &Integrated & Region 1 & Region 2 & Region 3\\ 
\hline
Te(O\,{\sc iii}) K                                      &19442$\pm$443 &19925$\pm$635 &20665$\pm$584 &19951$\pm$333\\
Ne(S\,{\sc ii}) cm$^{-3}$                               & $\sim$100    & $\sim$100    & $\sim$100    & $\sim$100\\ 
O$^{+}$/H$^{+} \times$10$^{5}$                          & 0.51$\pm$0.02& 0.37$\pm$0.02 & 0.22$\pm$0.01& 0.21$\pm$0.01\\ 
O$^{++}$/H$^{+} \times$10$^{5}$                         & 2.92$\pm$0.20& 2.97$\pm$0.28 & 2.18$\pm$0.17& 3.05$\pm$0.15 \\ 
O/H $\times$10$^{5}$                                    & 3.43$\pm$0.22& 3.35$\pm$0.30 & 2.40$\pm$0.19& 3.26$\pm$0.15 \\ 
12+log(O/H)                                             & 7.53$\pm$0.06& 7.52$\pm$0.09 & 7.38$\pm$0.08& 7.51$\pm$0.05  \\ 
N$^{+}$/H$^{+} \times$10$^{6}$                          & 0.14$\pm$0.01& 0.09$\pm$0.01 & 0.15$\pm$0.01& 0.06$\pm$0.01 \\ 
ICF(N)                                                  & 6.77$\pm$0.69& 9.02$\pm$1.28 &10.90$\pm$1.34&15.20$\pm$1.14\\ 
N/H $\times$10$^{6}$                                    & 0.93$\pm$0.12& 0.86$\pm$0.15 & 1.65$\pm$0.26& 0.92$\pm$0.08 \\ 
12+log(N/H)                                             & 5.97$\pm$0.13& 5.93$\pm$0.18 & 6.22$\pm$0.15& 5.96$\pm$0.09\\ 
log(N/O)                                                &-1.57$\pm$0.19&-1.59$\pm$0.27 &-1.16$\pm$0.23&-1.55$\pm$0.14\\ 
\hline
He\,{\sc ii} $\lambda$4686/H$\beta$                     & $\cdots$   & $\cdots$ & $\cdots$ &0.0122$\pm$0.0029\\ 
log($\left[OIII\right]\lambda$5007/H$\beta$)            & 0.70$\pm$0.03 &0.73$\pm$0.02 &0.63$\pm$0.03   &0.74$\pm$0.05\\ 
log($\left[NII\right]\lambda$6584/H$\alpha$)            &-2.19$\pm$0.12 &-2.34$\pm$0.12 &-2.13$\pm$0.11 &-2.54$\pm$0.06\\ 
log($\left[SII\right]\lambda\lambda$6717,6731/H$\alpha$)&-1.66$\pm$0.16 &-1.77$\pm$0.14 &-1.55$\pm$0.12 &-2.11$\pm$0.12 \\ 
log($\left[OI\right]\lambda$6300/H$\alpha$)             &-2.48$\pm$0.30 &-2.46$\pm$0.18 &-2.30$\pm$0.14 &-2.75$\pm$0.08\\ 

\hline
\end{tabular}
\label{integrated_properties_1}
\end{minipage}
\end{table*}

% ===================== Table 6 ====================================================================
\begin{table}
 \centering
 \begin{minipage}{100mm}
  \caption{Comparison of the chemical properties of regions 1 and 3.}
  \begin{tabular}{@{}lccccclrlr@{}}
  \hline      
&    \multicolumn{3}{c}{Region 1}& \multicolumn{3}{c}{Region 3}\\
& Mean & SDEV\footnote{Standard deviation of the spaxels.} & 
$\mid\Delta$\footnote{Difference between minimum and maximum values.}$\mid$& Mean 
& SDEV & $\mid\Delta\mid$\\
 \hline
O$^{+}$/H$^{+}$ ($\times$ 10$^{5}$) & 0.36&0.06&0.19& 0.23&0.05&0.15\\
O$^{++}$/H$^{+}$ ($\times$ 10$^{5}$)& 2.83&0.31&0.90& 3.37&1.09&3.16\\
12 + log (O/H)                      & 7.50&0.04&0.11& 7.54&0.14&0.41\\
N$^{+}$/H$^{+}$ ($\times$ 10$^{6}$) & 0.09&0.02&0.06& 0.05&0.02&0.05\\
12 + log (N/H)                      & 5.90&0.19&0.46& 5.86&0.19&0.63\\
log (N/O)                           &-1.60&0.16&0.39&-1.68&0.10&0.38\\ 
\hline
\end{tabular}
\label{table_static}
\end{minipage}
\end{table}

The oxygen abundance in the galaxy varies from 7.31 to 7.72 (see Figure \ref{abundance_maps}).
The integrated and mean values are almost equal for regions 1 and 3, with values of 
12+log(O/H)=7.52$\pm$0.09 and 7.51$\pm$0.05, respectively.
While, in region 2, we found an oxygen abundance of 12+log(O/H)=7.38$\pm$0.08.
The integrated abundance 12+log(O/H)=7.53$\pm$0.06 determined here agrees, within 
the errors, with the value reported by Thuan \& Izotov (\cite{TI05}) of 7.473$\pm$0.012. 
Our oxygen abundance determination for region 1, 2 and 3 agrees, within the errors, with the ones found by 
Izotov \& Thuan (\cite{IT07}) with 12+log(O/H)=7.450$\pm$0.012 and 7.562$\pm$0.013, 
for their regions 1 and 2, respectively.  
With regard to the nitrogen abundance, we derive an integrated value of 
12+log(N/H)=5.97$\pm$0.13, which is consistent, within the errors, with the values in individual regions 
(5.93$\pm$0.18, 6.22$\pm$0.15 and 5.96$\pm$0.09, for regions 1, 2 and 3, respectively).
The log(N/O) ratio (Fig.~\ref{abundance_maps}) ranges from -1.85  to -1.39  
with a mean value of -1.60$\pm$0.16 and -1.68$\pm$0.10 in regions 1 and 3, respectively. 
Finally, we found an integrated value of log(N/O)=-1.57$\pm$0.19 
and log(N/O) = -1.59$\pm$0.27 and -1.55$\pm$0.14 for regions 1 and 3, 
while the region 2 has a value of log(N/O)=-1.16$\pm$0.23, respectively.

\begin{figure}
\centering
\includegraphics[width=90mm]{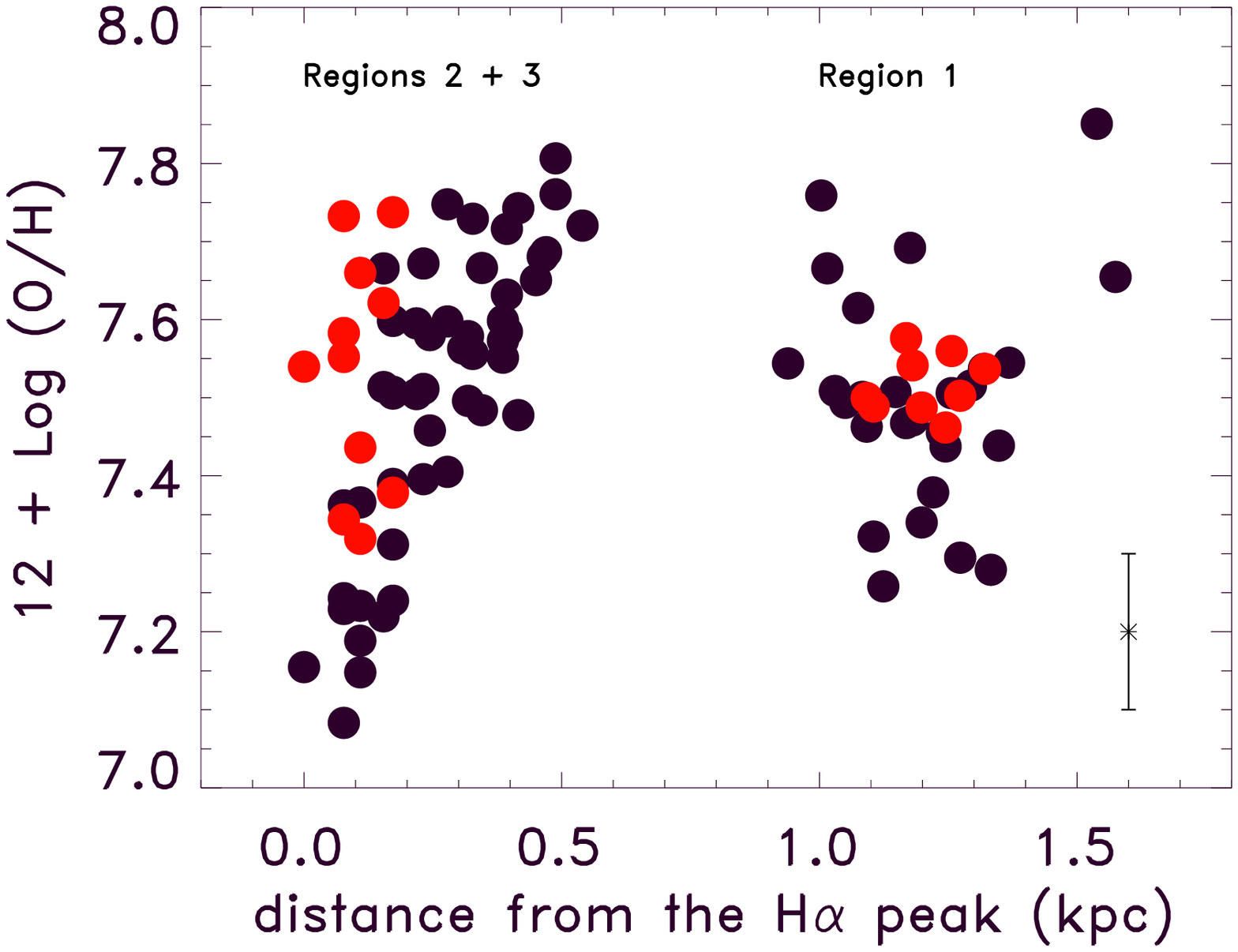}
 \caption{12+log(O/H) radial distribution in HS\ 2236+1344.
Red circles correspond to the data points obtained using the T$_{e}$(O\,{\sc iii}) and 
black circles to data points obtained using the Denicol\'o et al. (\cite{D02}) calibration.
The mean error is shown in the lower of the figure.
}
\label{abundance_dist}
\end{figure}

Figure \ref{abundance_dist} shows the radial distribution of the 
oxygen abundance with respect to the peak of H$\alpha$ emission in the 
spaxels (red data points) where the [O\,{\sc iii}] $\lambda$4363 emission line was detected. 
We also include oxygen abundance determinations based on empirical calibrations
(black data points in Figure \ref{abundance_dist}) for spaxels 
where a reliable determination of T$_e$(O\,{\sc iii}) was not possible. 
In this case, the O/H abundance was derived by applying the relation between the line ratio of 
[N\,{\sc ii}]$\lambda$6584/H$\alpha$ with the oxygen abundance from Denicol\'o et al. (\cite{D02}), i.e. 
12+log(O/H)=9.12($\pm$0.05)+0.73($\pm$0.10)$\times$N2, with N2=log([N\,{\sc ii}] $\lambda$6584/H$\alpha$).
The average oxygen abundance based on T$_e$(O\,{\sc iii}) was determined to be
12+log(O/H)=7.50 and 7.54 for regions 1 and 3, respectively, slightly above the value from the 
Denicol\'o's relation ($\sim$7.48), having a typical uncertainty of 0.18 dex.
However, the values based on the empirical calibration are, in some cases, by $\sim$0.4 dex lower 
than those relying on T$_{e}$(O\,{\sc iii}), at least in the peak of H$\alpha$ emission. 
Therefore, this empirical calibration method provides a useful mean value for the oxygen abundance 
in the ISM but is less suited for a detailed study of possible spatial variations 
(e.g., L\'opez-S\'anchez et al. \cite{LS11}).

% ===================== Figure 12 ====================================================================

\begin{figure}
\centering
\includegraphics[width=95mm]{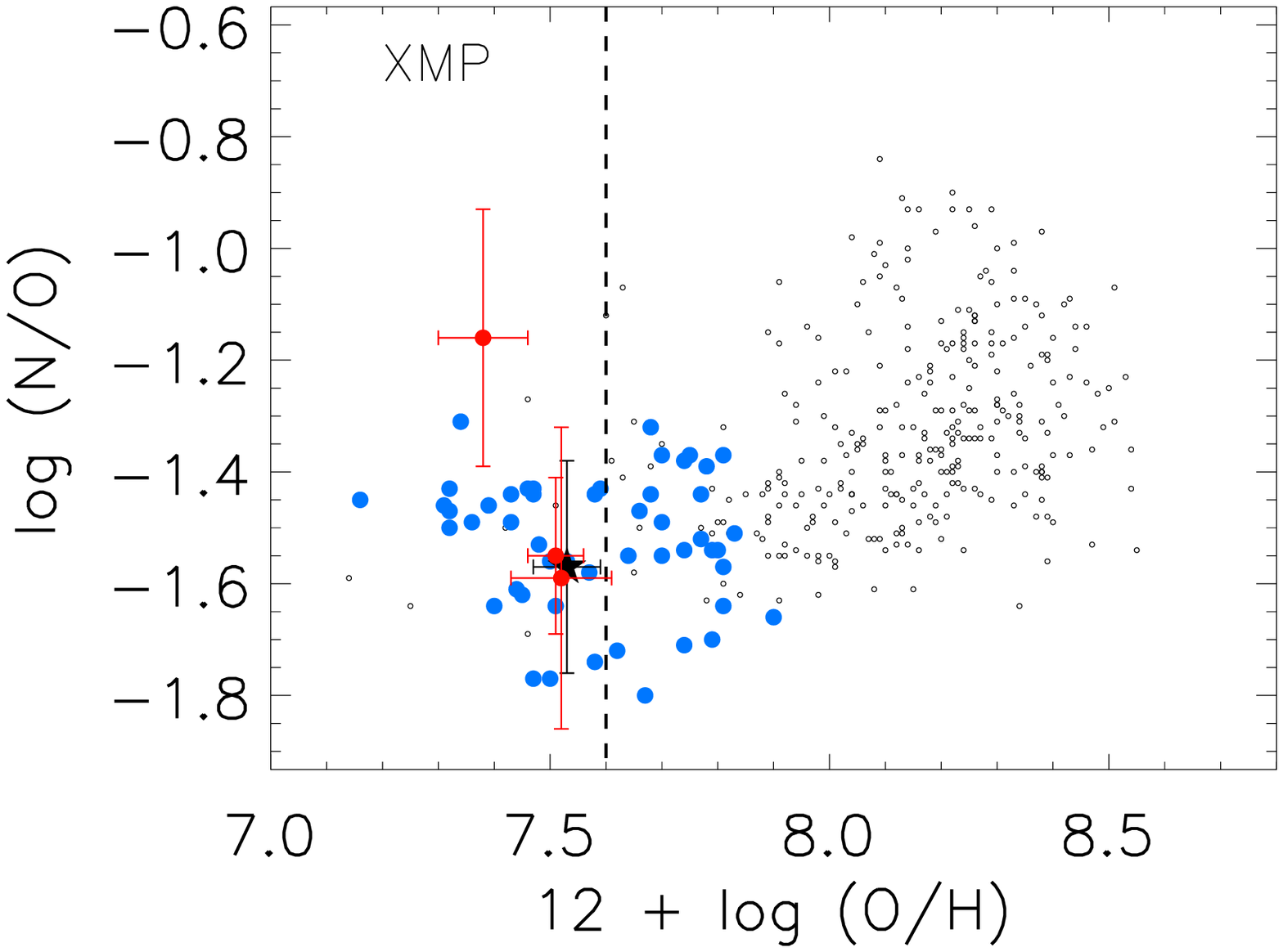}
 \caption{log(N/O) vs. 12+log(O/H) abundances. The values of the three GH\,{\sc ii}Rs resolved in
HS\ 2236+1344 are plotted by red points. The black star corresponds with the integrated value
of the galaxy. While the small black open dots corresponds with values found by Izotov et al. (\cite{I06b})
for a large sample of SDSS starburst galaxies and blue symbols show data from Izotov et al. (\cite{I12}).
}
\label{fig_NO_O}
\end{figure}

In Fig.~\ref{fig_NO_O} we show the relation between log(N/O) and 12+log(O/H), which includes
the values of the three GH\,{\sc ii}Rs resolved in HS\ 2236+1344 (red points). 
In the same figure, the black star indicates the integrated value of our analyzed galaxy. 
The small black open dots correspond to values published by Izotov et al. (\cite{I06b})
for a large sample of SDSS starburst galaxies, while
the blue data points are values from Izotov et al. (\cite{I12}) for a sample of XBCDs.
It is interesting to note that the integrated N/O value of the galaxy 
is similar, within the errors, to the ones obtained in regions 1, 2 and 3, and those of other BCD
galaxies of similar metallicity (see Fig.~\ref{fig_NO_O}).
Indeed, the O/H and N/O abundances from the integrated spectrum is in
excellent agreement, within the errors, with the mean value of the spaxels for regions 1 and 3.
Therefore, our results appear normal for XBCDs, which are characterized by a plateau at log(N/O)$\sim$-1.6 
(e.g., Edmunds \& Pagel \cite{EP78}, Alloin et al. \cite{A79}, Izotov \& Thuan \cite{IT99}). 

% ===================== Figure 13 ====================================================================

XBCDs are found to fall off the mass/luminosity -- metallicity relation compiled from literature data 
in Filho et al. (\cite{Filho13}), which is consistent with the presence of chemically unpolluted gas in these systems.
It is worth remarking in this context that S\'anchez almeida et al. (\cite{S14}) studied, 
using long-slit spectroscopy, the oxygen abundance along the major axis
of seven SF galaxies, including HS 2236+1344. In that study,
they find that the latter shows a central metallicity decrease by $\sim$0.5 dex, 
which they ascribe to accretion of metal-poor gas from the halo.
It is long known (e.g., Thuan et al. \cite{T05}) that the gas in the halos of H\,{\sc ii}/BCD galaxies is very metal
poor, a fact making infall and mixing with gas from the halo
a plausible explanation for the reported local decrease in the
metallicity of the ionized gas phase.
From the present data, however, given the uncertainties associated with
our determinations of O and N, we consider these chemical species to be
well mixed and homogeneously distributed over the ISM of the galaxy (e.g., Lagos et al. \cite{L09,L12}), albeit 
a slightly, and uncertain, decreased oxygen abundance was found in the faintest GH\,{\sc ii}R studied
(12+log(O/H)=7.38$\pm$0.08). 

% %%%%%%%%%%%%%%%%%%%%%%%%%%%%%%%%%%%%%%%%%%%%%%%%%%%%%%%%%%%%%%%%%%%%%%%%%%
\subsection{The surface brightness profile of HS\ 2236+1344}\label{morphology}
% %%%%%%%%%%%%%%%%%%%%%%%%%%%%%%%%%%%%%%%%%%%%%%%%%%%%%%%%%%%%%%%%%%%%%%%%%%

The surface brightness profile of HS\ 2236+1344 agrees with the detection of an underlying more evolved 
stellar host with a luminosity-weighted age of 0.5\dots2 Gyr in most XBCDs studied as yet 
(Papaderos et al. \cite{P08}), even though the estimated ages will depend, besides 
the poorly constrained SFH, also on the importance of radial stellar migration and the associated \emph{radial 
mass filtering effect} described in Papaderos et al. (\cite{Papaderos02}). 

\begin{figure}
\centering
\includegraphics[width=78mm]{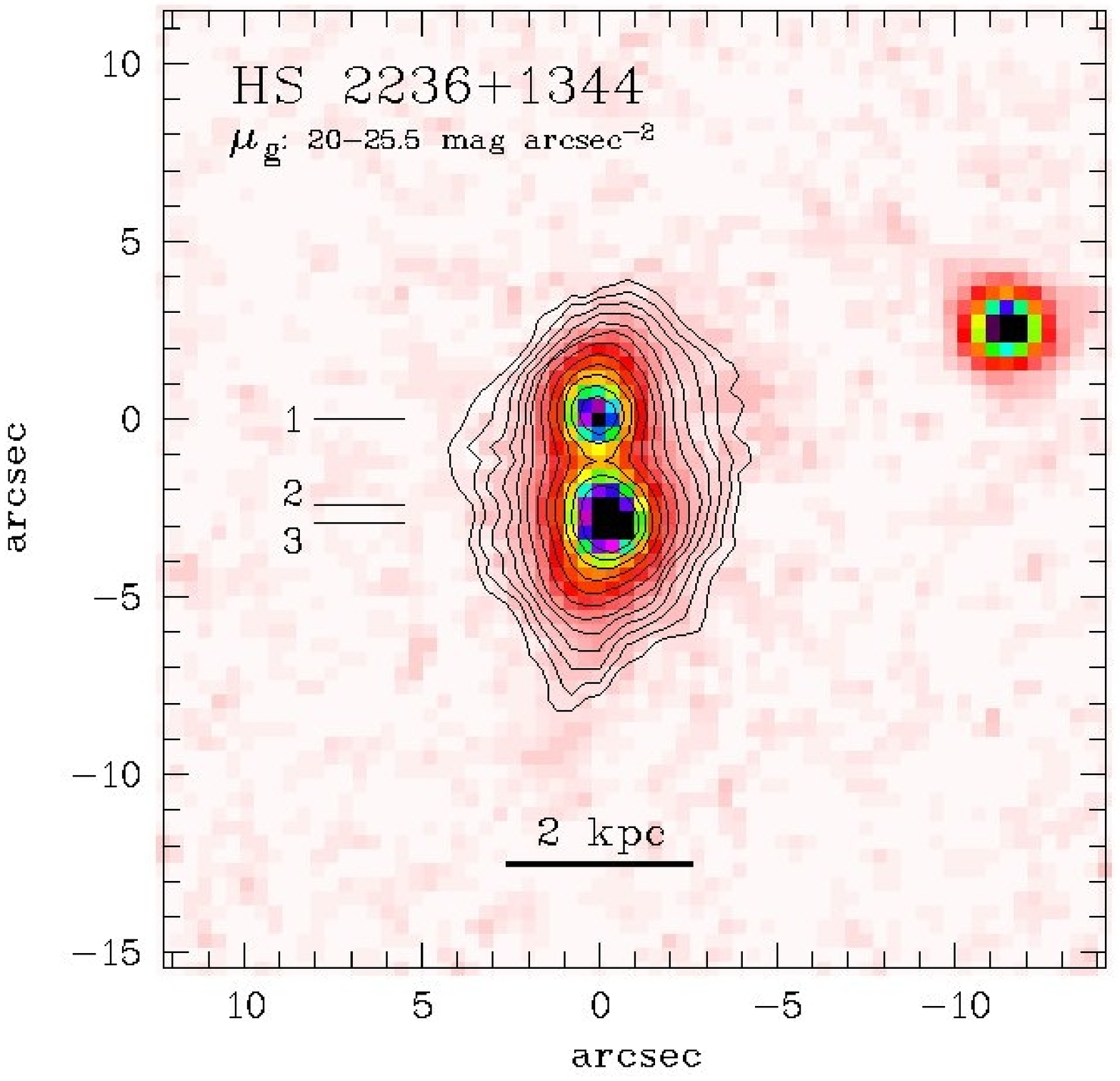}
 \caption{SDSS $g$~band image of HS\ 2236+1344 with overlaid contours between 20 and 25.5 $g$~mag~arcsec$^{-2}$
 in steps of 0.5 mag. The three GH\,{\sc ii}Rs in the XBCD are labeled.}
\label{contours}
\end{figure}

\begin{figure}
\centering
\includegraphics[width=77.8mm,angle=270]{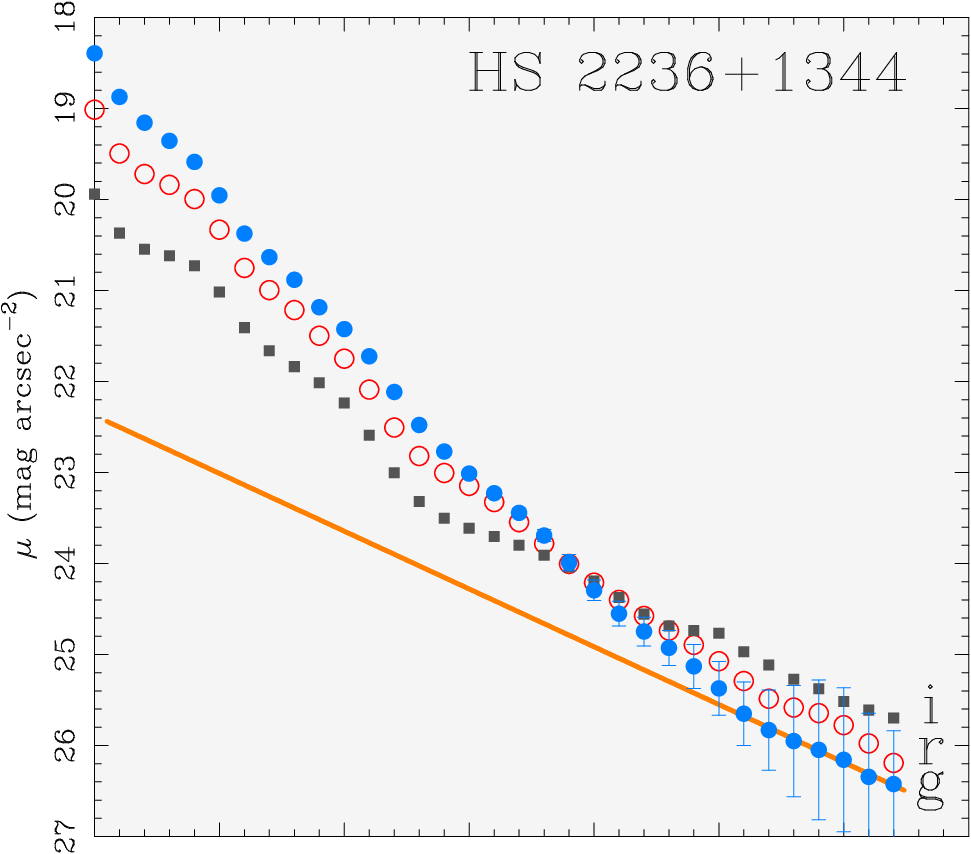}
\includegraphics[width=42.9mm,angle=270]{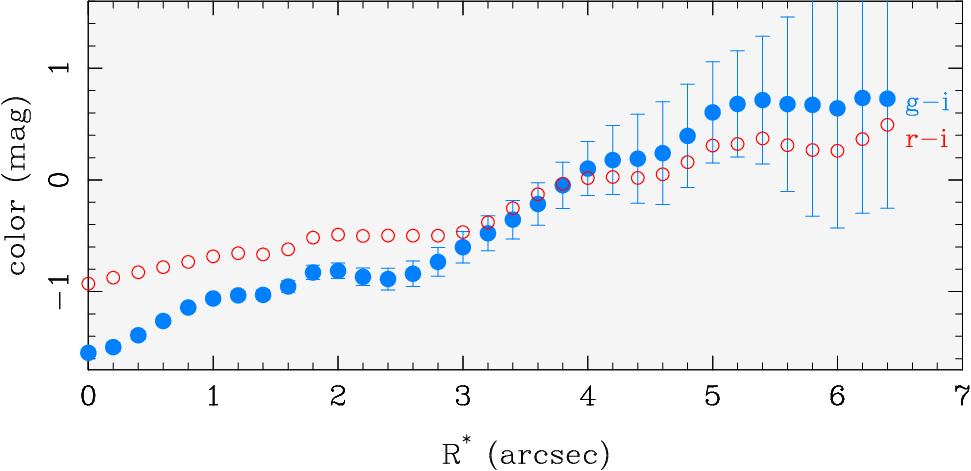}
 \caption{SDSS $g$, $r$ and $i$ band surface brightness profiles (upper panel) and 
$g$--$i$ and $r$--$i$ color profiles (lower panel) of HS\ 2236+1344.} 
\label{sbps}
\end{figure}

From the SDSS $g$-band contour map in Fig.~\ref{contours} it is apparent that the three GH\,{\sc ii}Rs in 
HS\ 2236+1344 are hosted by a more extended underlying LSB component with a moderately smooth
morphology. This LSB host dominates the light for surface brightness levels $\mu_g \ga 24$ mag arcsec$^{-2}$
and can be studied using SDSS data down to $\mu_g \simeq 25.5$ mag arcsec$^{-2}$.
The surface brightness profiles of the XBCD in $g$, $r$ and $i$ (Fig. \ref{sbps}, upper panel) were derived with 
method~iv by Papaderos et al. (\cite{Papaderos02}) (also referred to as LAZY by Noeske et al. \cite{Noeske06}). 
They show the typically complex radial intensity distribution of BCDs/XBCDs which is characterized
by an outer exponential component, corresponding to the LSB host, and a steep luminosity increase at smaller radii 
that is due to the young stellar population and nebular emission in the starburst component 
(e.g., Papaderos et al. \cite{P96a}; hereafter P96a). 
A fit to the $g$ band profile for $R^{\star}\geq$5\arcsec, i.e. beyond the 
\emph{transition radius} (P96a), where color profiles of BCDs level off to a nearly constant value, yields 
an exponential scale length of 1\farcs7$\pm$0\farcs1 ($\simeq$660$\pm$40 pc) and a total absolute magnitude 
of --15.3~mag for the LSB host. The latter contributes only about 15\% of the total luminosity, i.e. significantly 
less than the average value for normal-metallicity BCDs ($\sim$50\%; Papaderos et al. \cite{P96b}). 
With regard to its effective radius $r_{\rm eff}$ of 2\farcs2 ($\simeq$860~pc), HS\ 2236+1344 does not appear 
exceptional among BCDs/XBCDs.

The very blue $g$--$i$ color (--1.4 mag \dots --0.3 mag) of the galaxy in its central part 
(out to $\sim$1.5$r_{\rm eff}$) is only reproducible by a strong nebular emission contribution in the visual 
passband by the [O\,{\sc iii}] $\lambda\lambda$4959,5007 forbidden lines (Papaderos et al. \cite{P98}). 
This is in good agreement with the large measured H$\beta$ EWs (230 \dots 300 {\rm $\AA$}; Table~4) 
and previous spectroscopic and evolutionary synthesis studies by Guseva et al. (\cite{Guseva07}) 
which clearly reveal a strong nebular continuum contribution in the region around the Balmer jump.
  
The mean $g$--$i$ color in the LSB component of HS\ 2236+1344 (of $\sim 0.5$ mag; Fig.~\ref{sbps}, lower panel) 
corresponds to an age between $\sim$1 and $\sim$3.3 Gyr for a SFH approximated, respectively, by an instantaneous 
burst and an exponentially decreasing star formation rate (SFR) with an e-folding time of 1~Gyr 
(cf, e.g. Fig.~6 of Papaderos et al. \cite{P08}).
Judging from our IFU data, extended ionized gas emission, in this galaxy, does not dominate 
in the LSB periphery of HS 2236+1344, contrary to the case of I\ Zw\ 18 
(Papaderos et al. \cite{Papaderos02}, Papaderos \& \"{O}stlin \cite{PapaderosOstlin12}), 
thus the broadband colours of the LSB need not be corrected for this effect.

If the exponential LSB host of BCDs/XBCDs forms in an inside-out manner through radial migration of stars, then the 
associated \emph{stellar mass filtering} effect (Papaderos et al. \cite{Papaderos02}) 
will result into a stellar age overestimate, if colors are interpreted in terms of extended SFHs. 
In fact, a consequence of galaxy build-up through stellar 
migration is that instantaneous star formation models yield a better approximation to the true stellar age of the 
LSB host than the usually adopted continuous star formation SFH models 
(see for an update and a further discussion Papaderos \& \"{O}stlin \cite{PapaderosOstlin12}). 

% ===================== Figure 14 ====================================================================

% ===================== Figure 15 ====================================================================

\subsection{Star formation and the age of the current burst in HS\ 2236+1344}\label{sf_io_mech}
The H$\alpha$ flux in Table \ref{tb_lines_integrated} translate by the adopted distance 
and the Kennicutt (\cite{K98}) conversion formula after correction for a Kroupa IMF  (Calzetti et al. \cite{C07}),
SFR(M$_{\odot}$ yr$^{-1}$) = 5.3 $\times$ 10$^{-42}$ L(H$\alpha$) ergs s$^{-1}$, to an integrated SFR of 
0.587 M$_{\odot}$ yr$^{-1}$.
The SFRs in regions 1, 2 and 3 are estimated to be $\simeq$ 0.075, 0.124 and 0.216 M$_{\odot}$ yr$^{-1}$, 
respectively. The corresponding SFR per unit of area ($\Sigma_{SFR}$) is then $\sim$0.629, $\sim$0.695 and 
$\sim$1.810 M$_{\odot}$ yr$^{-1}$ kpc$^{-2}$, by some factors larger than the integrated value for the galaxy 
($\sim$0.145 M$_{\odot}$ yr$^{-1}$ kpc$^{-2}$). Note that the $\Sigma_{SFR}$ in individual regions 
are comparable with values determined for higher-metallicity starburst galaxies 
($>$0.1 M$_{\odot}$ yr$^{-1}$ kpc$^{-2}$; Daddi et al. \cite{D10}) and larger than typical values for normal spirals 
($\Sigma_{SFR}<$0.1 M$_{\odot}$ yr$^{-1}$ kpc$^{-2}$; Kennicutt \cite{K98}).
The integrated $\Sigma_{SFR}$ in HS\ 2236+1344 is also larger than the ones found in other 
BCD galaxies, such as Mrk\ 36 and UM\ 461 0.039 and 0.052 M$_{\odot}$ yr$^{-1}$ kpc$^{-2}$ 
(Lagos et al. \cite{L11}), respectively, and local cometary/tadpole SF galaxies 
($\sim$0.01 M$_{\odot}$ yr$^{-1}$ kpc$^{-2}$; Elmegreen et al. \cite{ED12}), yet comparable 
with the SFR in 30~Doradus (0.36 M$_{\odot}$ yr$^{-1}$ kpc$^{-2}$; Chen et al. \cite{C05}).

\begin{figure*}
\centering
\includegraphics[width=70mm]{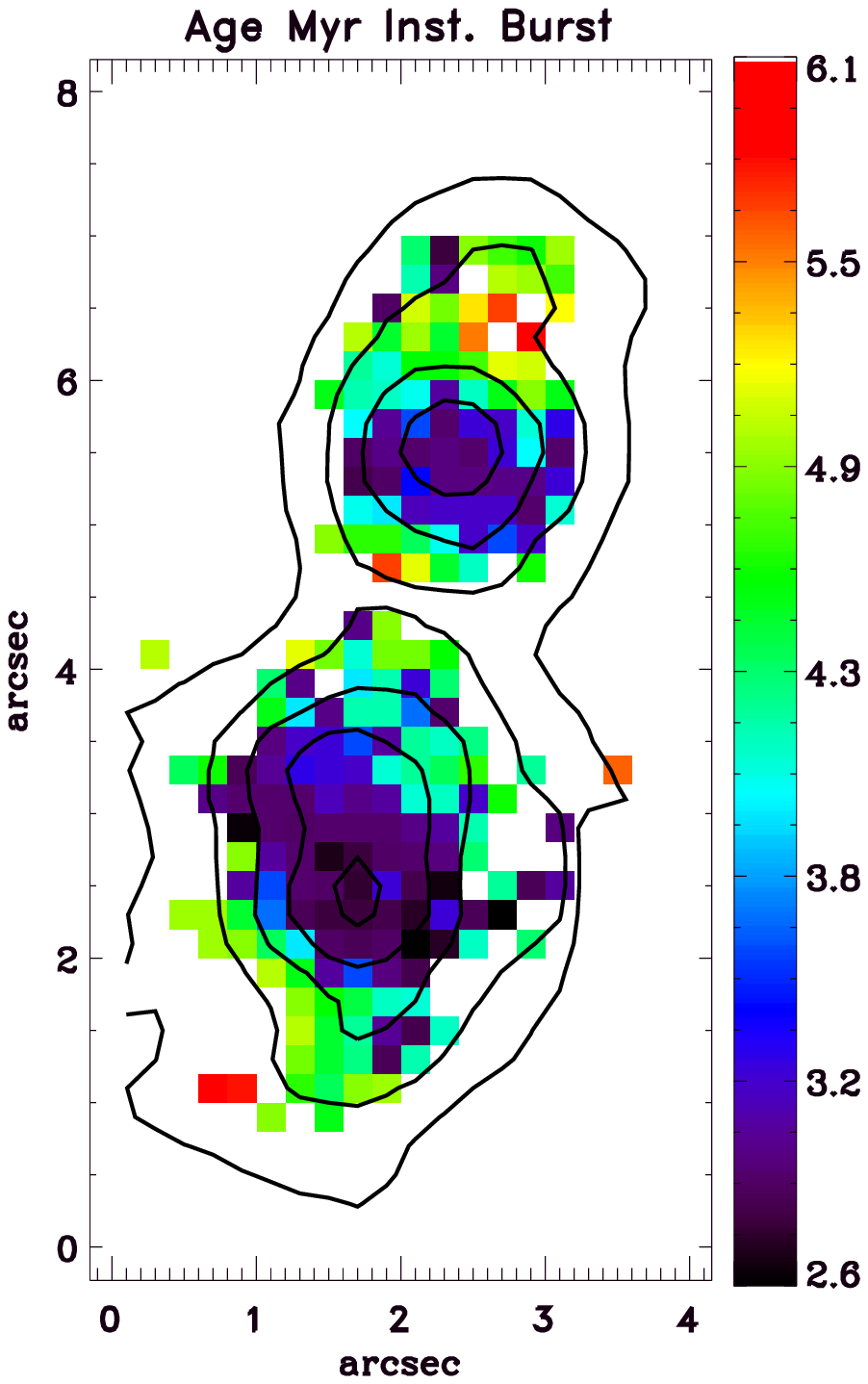}
\includegraphics[width=70mm]{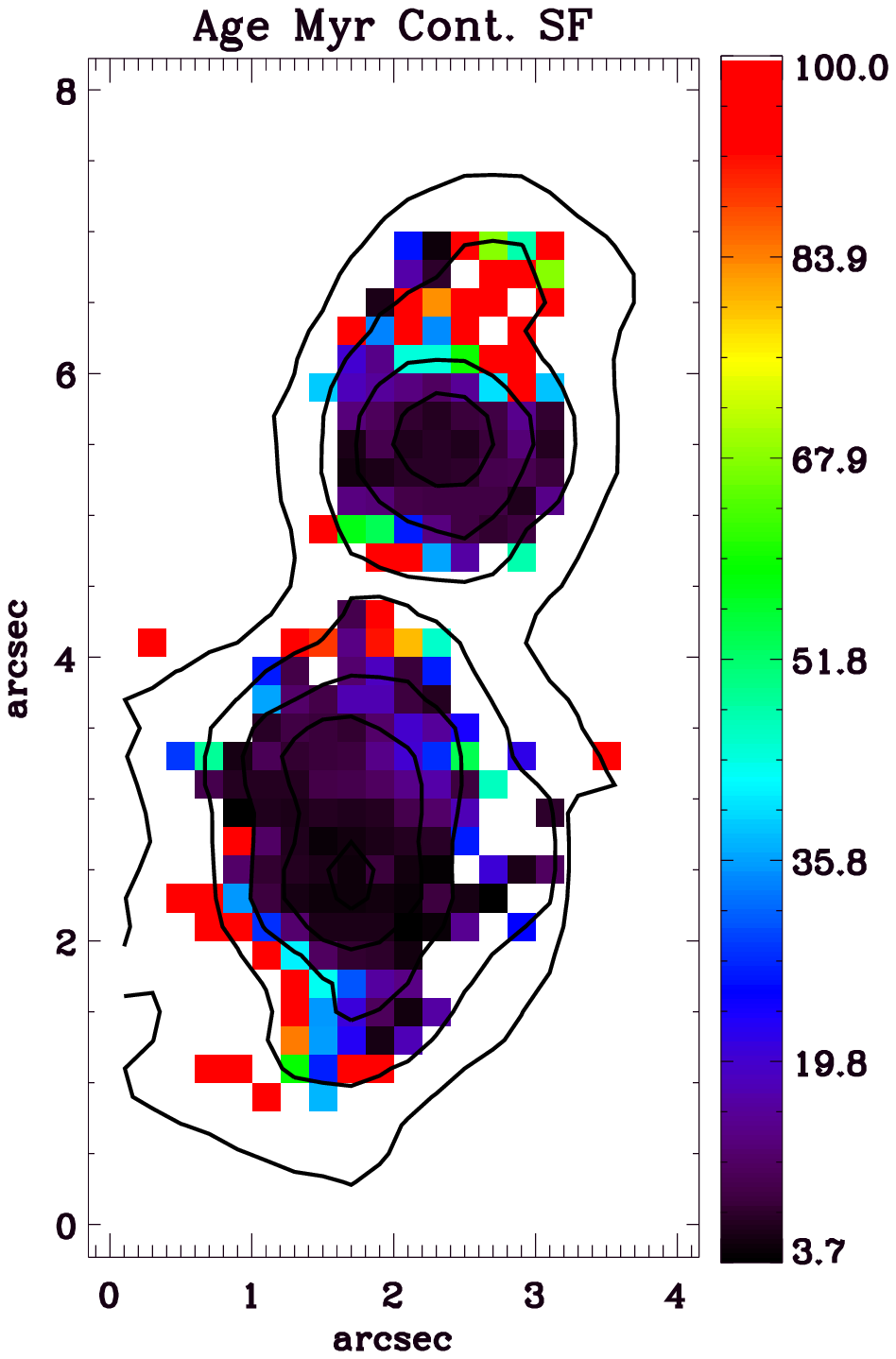}
 \caption{Age distribution in Myr obtained by direct conversion of EW(H$\alpha$) and EW(H$\beta$) maps into stellar
 ages using Starburst99 model predictions for an instantaneous burst (left panel) and continuous SF (right panel).
Overlaid are the H$\alpha$ flux contours. 
}
\label{age_dist}
\end{figure*}

Following common practice, an estimate on the starburst age in HS\ 2236+1344 can be obtained from comparison of the 
observed EW(H$\alpha$) and EW(H$\beta$) distribution with predictions from zero-dimensional 
evolutionary synthesis models. 
For this, we use Starburst99 models (Leitherer et al. \cite{L99}) for a metallicity Z=0.004, Geneva evolutionary 
stellar tracks, and a Kroupa initial mass function ($\varpropto$M$^{-1}$) with $\alpha$=1.3 for stellar masses 
between 0.1 to 0.5M$_{\odot}$ and $\alpha$=2.3 between 0.5 and 100M$_{\odot}$. 
For the SFH we consider the limiting cases of a single burst and continuous star formation 
with a constant SFR of 0.075 and 0.170 M$_{\odot}$yr$^{-1}$ for regions 1 and 2+3, 
respectively. The stellar age, in the FoV, obtained from Starburst99 models 
for these two SFHs (Fig.~\ref{age_dist}) is very low ($\sim$3 \dots\ $\sim$6 Myr) for the instantaneous burst model, 
and between $\sim$4 and $\sim$100 Myr for continuous star formation.
We note that the assumed metallicity (Z=0.004) does not significantly affect the derived age, given that we obtain
nearly the same age pattern when we assume a metallicity of Z=0.001. 
Figure~\ref{age_dist} suggests that regions 1 and 2+3 were formed almost coevally some 3 Myr ago, 
assuming an instantaneous burst, whereas their surroundings and the arm-like features appear to be slightly older.
%Evidently, the estimates above are indicative only, and are merely provided for the sake of comparison with 
%other studies of local SF galaxies. 
It should be born in mind that the EW maximum of nebular emission 
lines in starburst galaxies does not necessarily spatially coincide with the location of ionizing YSCs 
(e.g., Papaderos et al. \cite{P98,Papaderos02}, Guseva et al. \cite{Guseva04}, Lagos et al. \cite{L07}, 
Papaderos \& \"{O}stlin \cite{PapaderosOstlin12}), consequently EW maps are not always convertible 
into stellar age maps via standard evolutionary synthesis models (e.g., Starburst99).  
The unknown SFH is another important source of uncertainty in converting H$\alpha$ 
luminosities into SFRs for starburst galaxies (e.g., Weilbacher \& Fritze von Alvensleben \cite{Weilbacher01}), 
just like the fraction of ionizing Lyman continuum photons leaking out of SF regions
and galaxies (e.g., see also Guseva et al. \cite{Guseva07}, Rela\~no et al. \cite{Relano12}, 
Papaderos et al. \cite{P13}, Bergvall et al. \cite{Bergvall13}).
Therefore, the estimates above are indicative only, and are merely provided for the sake of comparison with 
other studies of local SF galaxies.

% ===================== Figure 16 ====================================================================

We re-iterate that several studies point to similar physical conditions across the ISM of XBCD (P12),  
H\,{\sc ii}/BCDs (e.g., UM 408, and \object{Tol 0104-388} and Tol 2146-391; Lagos et al. \cite{L09,L12}) and 
to nearly coeval SF activity in these systems on spatial scales of $\sim$1~kpc (e.g., Lagos et al. \cite{L11}).
Whereas most low-luminosity BCDs show a smooth morphology in their LSB host, luminous ($M_B\la -19$ mag) ones
often show tidal distortions on deep images (e.g., Bergvall \& \"{O}stlin \cite{BO02}, Lagos et al. \cite{L07}), 
suggesting strong interactions or merging as the primary triggering agent of their starburst activity
(e.g., Taylor et al. \cite{Taylor95}, Pustilnik et al. \cite{Pustilnik01}, 
Brosch et al. \cite{B04}, Ekta et al. \cite{E08}). 
If SF activity in HS\ 2236+1344 is due to a recent or ongoing merger, as suggested by Moiseev et al. (\cite{M10}), 
a flatter metallicity distribution may be evolve through transport and mixing of metals
(e.g., Rupke et al. \cite{R10}, Montuori et al. \cite{Mo10}) following tidal tail formation.
Thus far, in most studies on chemical abundances in SF dwarf galaxies (e.g., Lee \& Skillman \cite{LS04}, 
Kehrig \cite{K08}, Lagos et al. \cite{L09,L12}) there are no clear evidences of abundance variations. 
Therefore, the overall chemical homogeneity of the warm ISM in HS\ 2236+1344 suggests an efficient dispersal 
and mixing of heavy elements. In any case, we emphasize that given the uncertainties, in this study, 
we must consider the abundances across the galaxy as fairly homogeneous. 
On the other hand,  the estimated age of the GH\,{\sc ii}Rs, found in this galaxy, 
suggests that the current, large-scale, burst started recently and likely simultaneously.
Therefore, the triggering mechanism (i.e., minor interactions or infall of gas from the halo)
may be related to the overall physical conditions of the ISM, 
particularly the gas surface density, in conjunction with small stochastic effects. 

Clearly, detailed studies of the ISM with IFU spectroscopy  
(see, e.g., Lagos \& Papaderos \cite{LP13}) have the potential of greatly improving our understanding 
of the evolution of chemical abundance patterns and star formation in H\,{\sc ii}/BCD galaxies.

\section{Summary}\label{summary}

Our spatially resolved study of the warm ISM in the XBCD galaxy HS\ 2236+1344 by means of IFU spectroscopy 
yields the following conclusions:

\begin{enumerate}
      
\item HS\ 2236+1344 contains three GH\,{\sc ii}Rs with an H$\alpha$ luminosity in the range
$\sim$ 1 $\dots$ 4$\times$ 10$^{40}$ erg/s. The high-angular resolution acquisition image of the galaxy 
additionally shows some faint arm-like structures in the close vicinity of the northern and southern GH\,{\sc ii}R, 
which might be of tidal origin or due to expanding gas shells. 
The structure of the ionized gas, as traced both by the emission lines and the continuum are fairly similar.

\item  The H$\alpha$ velocity field v$_{r}$(H$\alpha$) of the galaxy shows a smooth 
gradient along its major axis, with a difference of about 80 km s$^{-1}$ between 
its receding northwestern and approaching southeastern half. The observed velocity range 
is comparable to values determined for other BCDs (e.g., van Zee et al. \cite{v01}).

\item A comparison of EW(H$\alpha$) and EW(H$\beta$) maps with predictions from the evolutionary synthesis code 
Starburst99 suggests that the three GH\,{\sc ii}Rs in HS\ 2236+1344 were formed almost coevally less than 
$\sim$3 Myr ago. The estimated SFR surface density of 
$\Sigma_{SFR_{int}}\sim$0.2 M$_{\odot}$ yr$^{-1}$ kpc$^{-2}$ is larger than the ones found
in other low-luminosity H\,{\sc ii}/BCD galaxies (Lagos et al. \cite{L11}) and local
cometary/tadpole SF galaxies (Elmegreen et al. 2012).
Surface brightness profiles, derived from SDSS data, reveal a compact underlying stellar host, which presumably 
indicates that HS\ 2236+1344 has undergone previous SF activity, as was found to be the case in 
most XMP BCDs studied as yet (e.g., Papaderos et al. \cite{P08}).

\item The high-ionization He\,{\sc ii} $\lambda$4686 emission line is detected in  
HS\ 2236+1344. Similar to many BCDs with He\,{\sc ii} $\lambda$4686 emission, HS\ 2236+1344 shows no WR bump.
The spatial distribution of this high-ionization emission, observed in this study, 
suggests that H\,{\sc ii} $\lambda$4686 in HS\ 2236+1344 is associated with the current burst of star formation.  

\item We calculated the spatial distribution of oxygen and nitrogen, and their ratio, 
based on spaxels with reliable T$_{e}$(O\,{\sc iii}) determinations.
The oxygen abundance in the three resolved GH\,{\sc ii}Rs is consistent, within uncertainties, 
with the integrated value of 12+log(O/H)=7.53$\pm$0.06, suggesting a nearly constant metallicity 
across the ISM  of HS\ 2236+1344. 
With regard to the nitrogen-to-oxygen abundance ratio, we derive a value of 
log(O/N)=-1.57$\pm$0.19 for the integrated spectrum, 
in good agreement with N/O determinations for other XBCDs (log(N/O)$\simeq$-1.6). 
The overall chemical homogeneity of the warm ISM in HS\ 2236+1344 suggests,
in line with previous studies, an efficient dispersal and mixing of heavy elements in the 
lowest-metallicity H\,{\sc ii}/BCD galaxies.
   
\end{enumerate}
  
\begin{acknowledgements}
We would like thank the anonymous referee for his/her
comments and suggestions which substantially improved
the paper. P.L. is supported by a Post-Doctoral grant SFRH/BPD/72308/2010, 
funded by Funda\c{c}\~{a}o para a Ci\^{e}ncia e a Tecnologia (FCT).
P.P. is supported by Ciencia 2008 Contract, funded by
FCT/MCTES (Portugal) and POPH/FSE (EC).
J.M.G. is supported by a Post-Doctoral grant SFRH/BPD/66958/2009, funded by FCT (Portugal).
P.L., P.P. and J.M.G. acknowledge support by the FCT under project 
FCOMP-01-0124-FEDER-029170 (Reference FCT PTDC/FIS-AST/3214/2012), funded by the FEDER program. 
A.V.S.C. acknowledge financial support from Consejo Nacional de Investigaciones Cient\'{\i}ficas y T\'ecnicas, 
Agencia Nacional de Promoci\'on Cient\'{\i}fica y Tecnol\'ogica (PICT 2010-0410), 
and Universidad Nacional de La Plata (Argentina).
This research has made use of the NASA/IPAC Extragalactic Database (NED) which is operated 
by the Jet Propulsion Laboratory, California Institute of Technology, under contract with 
the National Aeronautics and Space Administration. 
Based on observations obtained at the Gemini Observatory (Program ID: GN-2010B-Q-69), which 
is operated by the Association of Universities for Research in Astronomy, Inc., under a cooperative agreement 
with the NSF on behalf of the Gemini partnership: the National Science Foundation (United States), the Science 
and Technology Facilities Council (United Kingdom), the National Research Council (Canada), CONICYT (Chile), 
the Australian Research Council (Australia), Minist\'erio da Ci\^encia e Tecnologia (Brazil) and Ministerio de Ciencia,
Tecnolog\'{\i}a e Innovaci\'on Productiva (Argentina).

\end{acknowledgements}

\end{document}